\documentclass[11pt,a4paper]{article}

\usepackage[dvips]{graphicx}
\usepackage{amsfonts,amssymb,eucal,amsmath,epsfig}

\newtheorem{theorem}{Theorem}[section]
\newtheorem{example}[theorem]{Example}
\newtheorem{lemma}[theorem]{Lemma}
\newtheorem{proposition}[theorem]{Proposition}
\newtheorem{definition}{Definition}[section]
\newtheorem{remark}{Remark}[section]

\begin{document}

\begin{center}
\textbf{Functionals with values in the Non-Archimedean field of
Laurent series and their applications to the equations of
elasticity theory}

\emph{Dedicated to Tatsiana  Radyna}
\end{center}

\begin{center}
\textbf{Mikalai Radyna}\footnote{e-mail:~ mik\_ radyna@yahoo.com}
\medskip

{\it Institute of Mathematics

National Academy of Sciences of Belarus

Surganova 11,  Minsk, 220072

BELARUS

E-MAIL: kolya@im.bas-net.by}

\end{center}

\begin{abstract}
Functionals with values in Non-Archimedean field of Laurent series
applied to the definition of generalized solution (in the form of
soliton and shock wave) of the Hopf equation and equations  of
elasticity theory. Calculation method for the profile of
infinitely narrow soliton and  shock wave is proposed. Applying
this method, calculations of profiles are reduced to the nonlinear
system of algebraic equations in $\mathbf{R}^{n+1}$, $n>1$. It is
shown that there is a possibility to find out some of the
solutions of this system using the Newton iteration method.
Examples and numerical tests are considered.
\end{abstract}

\noindent KEY WORDS: generalized functions, distributions,
conservation law, Hopf equation, equations of elasticity theory,
soliton, shock wave
\medskip

\noindent MSC (2000): 46F99, 46F30, 35D99, 35L65, 74J35, 74J40,
76L05

\section{Introduction}
While working at Los Alamos in 1943-44, von Neumann became
convinced that the calculation of the flows of compressible fluids
containing strong shocks could be accomplished only by numerical
methods. He conceived the idea of capturing shocks, i.e., of
ignoring the presence of a discontinuity. Employing a Lagrangian
description of compressible flow, setting heat conduction and
viscosity equal to zero, von Neumann replaced space and time
derivatives by symmetric difference quotients. Calculations using
this scheme were carried out; the approximation resulting from
these calculations (see \cite{Neumann}) showed oscillations on the
mesh scale behind the shock. Von Neumann boldly conjectured that
the oscillations in velocity represent the heat energy created by
the irreversible action of the shock, and that as $\Delta x$ and
$\Delta t$ tend to zero, the approximate solutions tend in the
weak sense to the discontinuous solution of the equations of
compressible flow.

In \cite{Lax0} it was counterconjectured  that von Neumann was
wrong in his surmise, i.e., that although the approximate
solutions constructed by his method do converge weakly, the weak
limit fails to satisfy the law of conservation of energy.

In \cite{Goodman-Lax} J.Goodman and P.Lax investigated von
Neumann's algorithm applied to the scalar equation
\begin{equation}\label{HopfEquation}
u_t+uu_x=0
\end{equation}
(it is called the \emph{Hopf equation}), in the semidiscrete case.
Using numerical experimentation and analytical techniques the
demonstrated the weak convergence of the oscillatory
approximations, and that the weak limit fails to satisfy the
scalar equation in question.

Von Neumann's dream of capturing shocks was realized in his joint
work with Richtmyer in 1950, see \cite{Neumann-Richtmyer}.
Oscillations were eliminated by the judicious use of artificial
viscosity; solutions constructed by this method converge uniformly
except in a neighborhood of shocks, where they remain bounded and
are spread out over a few mesh intervals. The limits appear to
satisfy the conservation laws of compressible flow. The
conservation of mass and momentum is the consequence of having
approximated these equations by difference equations in
conservation form; but the von Neumann-Richtmyer difference
approximation to the energy equation is not in conservation form.

In the paper  \cite{Hou-Lax} T.Hou and P.Lax compared the results
of a von Neu\-mann-Richtmyer calculation with the weak limit of of
calculations performed by  von Neumann's original method. P.Lax in
the paper \cite{Hou-Lax} asserts that the difference scheme of von
Neumann, because of the centering of the difference quotients, is
\emph{dispersive}; it is this quality that is responsible for the
oscillatory nature of the solutions.

E.Hopf \cite{Hopf} studied  (\ref{HopfEquation}) defined the
generalized solution for the this equation. He considered the
approximating equation
\begin{equation}\label{BurgersEquation}
u_t+uu_x=\mu u_{xx}\,\, \textrm{where}\,\, \mu\to 0
\end{equation}
for the equation  (\ref{HopfEquation}).

By a generalized solution $u$ of (\ref{HopfEquation}) or
(\ref{BurgersEquation}), $\mu =0$ E.Hopf meant a function $u$ that
is measurable and quadratically integrable in every closed
rectangle in the open semiplane  $t>0$ and that satisfies the
relation
\begin{equation}\label{ClassicGSolution}
\iint \left[ ug_t+\frac{u^2}{2} g_x\right] dx dt =0,
\end{equation}
where $g$ is an arbitrary function of class $C^1$ in  $t>0$ that
vanishes outside some circle lying entirely in $t>0$. He assert:
Every limit function  $u$ obtained from the solution of
(\ref{BurgersEquation}) as $\mu\to +0$ is a generalized solution
of (\ref{HopfEquation}). By a generalized solution $u$ of
(\ref{BurgersEquation}) E.Hopf meant a function $u$ that is
measurable and quadratically integrable in every closed rectangle
in the open semiplane  $t>0$ and that satisfies the relation
\begin{equation}\label{ClassicGSolution1}
\iint \left[ uf_t+\frac{u^2}{2} f_x +\mu uf_{xx}\right] dx dt =0,
\end{equation}
where $f$ is an arbitrary function of class $C^2$ in  $t>0$ that
vanishes outside some circle lying entirely in $t>0$.

This method is called the ``disappearing viscosity'' method. It
was developed by E.Hopf \cite{Hopf}, O.A.Oleinik \cite{Oleinik},
P.Lax \cite{Lax}, \cite{Hou-Lax}.

There is also the ``zero dispersion limit'' method developed by
P.Lax \cite{Lax0}, \cite{Goodman-Lax}  Maslov V.P and his
collaborators \cite{Danilov-Maslov-Schelkovich},
\cite{Maslov-Omel'yanov}. The idea is to use the following
equation
\begin{equation}\label{KdVEquation}
u_t+uu_x=\varepsilon^2 u_{xxx},\,\, \varepsilon\to 0.
\end{equation}
for the (\ref{HopfEquation}). V.P.Maslov and his collaborators
constructed an asymptotic solutions for the (\ref{HopfEquation}).
They speculated on the fact that equations (\ref{BurgersEquation})
and (\ref{KdVEquation}) have solutions in the class $C^{\infty}$
functions. For example, a particular solution of the equation
(\ref{KdVEquation}) is the function
$$c+4-12\cosh^{-2}\left(\frac{x-ct}{\varepsilon}\right),\,\,
c>0,$$ which represent so-called an infinitely narrow soliton. A
particular solution of the (\ref{BurgersEquation}) will be, for
instance, $$2-2\tanh\left(\frac{x-2t}{\mu}\right)$$ which
converges to a discontinues function.

In the other hand, in the paper \cite{Sobolev}  by S.L.Sobolev
(1936) were introduced mathematical basics of the theory of
generalized functions, was developed the idea of a generalized
function as a functional and was introduced the concept of
generalized solutions of a \emph{linear differential equation}.
These generalized functions and generalized solutions were
developed by L.Schwartz \cite{Schwartz}.  However,
Sobolev-Schwartz distributions can not applied to nonlinear
differential equations. For example, to substitute  generalized
function $H(x-vt)$ (where H is the Heaviside function) into the
({\ref{HopfEquation}) one need to define the product of two
distributions $H$ and $H'$ for the term $uu_x$. However, in 1954
L.Schwartz showed  that it is impossible to introduce an associate
multiplication in the space of distributions.

Starting from 1982 in the works by J.-F. Colombeau
\cite{Colombeau}, M. Oberguggenberger \cite{Oberguggenberger}, H.
Bi\-agioni \cite{Biagioni}, E. Rozin\-ger \cite{Rosinger}, A.Y. Le
Roux \cite{Colombeau-LeRoux}, Yu. Egorov \cite{Egorov}, J.-A.
Marti \cite{Marti}, A. Delcroix, D.Scarpal\'{e}\-zos
\cite{Delcroix-Scarpalezos}, B. Keyfitz \cite{Keyfitz}, A.
Antonevich, Ya. Radyno \cite{Antonevich-Radyno-Radyno}, T. Todorov
\cite{Todorov}, S. Pilipo\-vi\'{c}
\cite{Nedeljkov-Pilipovic-Scarpalezos} and others, a new theory of
generalized functions is developed. Such functions form the
algebra and contain distributions.

In general, nonlinear \ generalized \ functions \ are \ classes of
equivalent  smooth functions. Clearly that, one should pay
attention to this approach in order to consider nonlinear
differential equations. However we are now in a position to
develop a new point of view on generalized functions and their
applications to nonlinear equations. Namely, it is  necessary to
use an integral nature  of a conservation law. Conservation laws
are integral expressions from physical point of view and  it is
natural to consider an integral form of conservation laws.
Moreover, we want to develop new point of view on conservation
laws using the concept of functionals with values in the
Non-Archimedean field of Laurent series. We call such functionals
as $\mathbf{R}\langle\varepsilon\rangle$-distributions
\cite{Radyno1}, \cite{Radyna}. In addition, we give the definition
of the special kind of solutions of the some conservation laws in
the sense of $\mathbf{R}\langle\varepsilon\rangle$-distributions
and consider the method for the numerical calculations of the
smooth shocks and soliton like solutions of the  Hopf equation and
equations of elasticity theory in the mentioned sense. This method
based on orthogonal system of the Hermite functions as a base for
calculation of such solutions (i.e. shocks and infinitely narrow
solitons). Calculations of profiles of infinitely narrow soliton
and shock wave are reduced  to the nonlinear system of algebraic
equations in $\mathbf{R}^{n+1}$, $n>1$. We proved, using the
Schauder fixed point theorem \cite{Schauder}, that the mentioned
system has at least one solution in $\mathbf{R}^{n+1}$. We showed
that there is possibility to find out some of the solutions of
this system using the Newton iteration method
\cite{Kantorovich-Akilov}. We considered examples and numerical
tests. We also should emphasis that proposed numerical approach do
not use a difference scheme.

First, let us consider a bit of theory which we will apply to
conservation laws.

\section{Non-Archimedean field of Laurent series and\\
$\mathbf{R}\langle\varepsilon\rangle$--distributions.}

The theory of Non-Archimedean \ fields \ was \ considered \ in the
book by A.H.Light\-stone and A.Robin\-son
\cite{Lightstone-Robinson}.

\begin{definition}\label{definition1}
A \emph{Laurent series} is a formal object
$$\sum_{n=0}^{\infty}
\xi_{n+k} \varepsilon^{n+k}$$ where $k$ is a fixed (i.e., fixed
for this Laurent series), each $\xi_i\in\mathbf{R}$, and either
$\xi_k\not= 0$ or each $\xi_i=0$.
\end{definition}

The Laurent series $\sum_{n=0}^{\infty} \xi_n \varepsilon^n $,
where $\xi_0=1$ and $\xi_n=0$ if $n>0$, is denoted by 1. It is
easy to see that the Laurent series is a field. Let us denote it
by $\mathbf{R}\langle\varepsilon\rangle$.  The norm on the field
of Laurent series can define $$|x|_\nu=e^{-\nu (x)}\quad
\textrm{for}\quad \textrm{each}\quad x\in
\mathbf{R}\langle\varepsilon\rangle $$ (in place of $e$ can use
any number greater then 1). The function $\nu(x)$ is a
Non-Archimedean valuation. Define $$ \nu(0)=\infty
\quad\textrm{and}\quad \nu\left(\sum_{n=0}^{\infty} \xi_{n+k}
\varepsilon^{n+k}\right)=k \quad\textrm{if}\quad
\sum_{n=0}^{\infty} \xi_{n+k} \varepsilon^{n+k}\not=0,\quad
\xi_k\not= 0.$$

The norm $|\cdot|_\nu$ have properties
\begin{enumerate}
\item $|x|_\nu=0 \quad\textrm{if and only if}\quad x=0,$

\item $|xy|_\nu=|x|_\nu\cdot |y|_\nu,$

\item $|x+y|_\nu\leq\max\left\{|x|_\nu,\, |y|_\nu\right\}.$
\end{enumerate}
Here, we propose a general construction of the
$\mathbf{R}\langle\varepsilon\rangle$--valued generalized
functions \cite{Radyno}. These objects are a natural
generalization of Sobolev-Schwartz distributions. We call them as
$\mathbf{R}\langle\varepsilon\rangle$--distributions.

\begin{enumerate}
\item Consider all functions $f(x,\varepsilon)\in
C^{\infty} (\mathbf{R}\times (0,1))$ such that integrals
$$\int\limits_{-\infty}^{+\infty} f(x,\varepsilon )\psi (x)dx$$
exist for any  $\varepsilon$ and for all $\psi (x)$  from a given
class of functions $\mathcal{X}$ ($\mathcal{X}$ can be $C^\infty_0
(\mathbf{R}),$ $\mathcal{S}(\mathbf{R})$ and etc.).

\item   Suppose also  that $\displaystyle\int\limits_{-\infty}^{+\infty}
f(x,\varepsilon )\psi (x)dx$ is a number $a_{f,\varepsilon}(\psi)$
from the field of Laurent series
$\mathbf{R}\langle\varepsilon\rangle$.

\item The two functions  $f(x,\varepsilon )$ and $g(x,\varepsilon )$
call equivalent with respect to test functions $\mathcal{X}$ if
and only if $$\int\limits_{-\infty}^{+\infty} f(x,\varepsilon
)\psi
(x)dx=a_{f,\varepsilon}(\psi)=a_{g,\varepsilon}(\psi)=\int\limits_{-\infty}^{+\infty}
g(x,\varepsilon )\psi (x)dx.$$ The equality means in sense of the
field of Laurent series $\mathbf{R}\langle\varepsilon\rangle$ for
all functions $\psi\in\mathcal{X}$. Classes of equivalent
functions call
$\mathbf{R}\langle\varepsilon\rangle$\emph{--functions}. The
expression $$\int\limits_{-\infty}^{+\infty} f(x,\varepsilon )\psi
(x)dx$$ associates a number from
$\mathbf{R}\langle\varepsilon\rangle$ with every $\psi$. Such a
quantity is called a functional. In this case a linear functional
map $\mathcal{X}$ into the Non-Archimedean field
$\mathbf{R}\langle\varepsilon\rangle$. Call these functionals as
$\mathbf{R}\langle\varepsilon\rangle$\emph{-distributions}.
\end{enumerate}
Thus,

\begin{proposition}
$\mathbf{R}\langle\varepsilon\rangle$--function
$f(x,\varepsilon)=0$ if and only if
$$\int\limits_{-\infty}^{+\infty} f(x,\varepsilon )\psi (x)dx=0\in
\mathbf{R}\langle\varepsilon\rangle$$ for every $\psi$ from
$\mathcal{X}$.

The set of all
$\mathbf{R}\langle\varepsilon\rangle$\emph{--distributions} denote
by $\mathcal{R}(\mathcal{X})$
\end{proposition}

\begin{remark}
Recall that the idea of representation \ of \ a \ function \ $f\in
L_{\it loc }^1(\mathbf{R})$ in terms of a linear functional
$$C_0^{\infty}(\mathbf{R})\ni\psi\longmapsto
\int\limits_{-\infty}^{+\infty} f(x)\psi (x)dx\in\mathbf{R} $$
based on well-known proposition that if  $f\in L_{\it loc
}^1(\mathbf{R})$ and $\displaystyle\int\limits_{-\infty}^{+\infty}
f(x)\psi (x)dx=0$ for any $\psi\in C_0^{\infty}(\mathbf{R})$ then
$f=0$ almost everywhere.
\end{remark}

Let us consider an example of the
$\mathbf{R}\langle\varepsilon\rangle$-distribution.

\begin{example}
Take $\mathcal{X}=C_0^{\infty}(\mathbf{R})$ and $f(x,\varepsilon
)=\varphi (x/\varepsilon)$,  $\varphi (x)\in
C_0^{\infty}(\mathbf{R})$ then
$\mathbf{R}\langle\varepsilon\rangle$--distribution can write in
the following form. $$\int\limits_{-\infty}^{+\infty} \varphi
(x/\varepsilon)\psi (x)dx=\varepsilon
\int\limits_{-\infty}^{+\infty} \varphi (x) dx \psi
(0)+\varepsilon^2 \displaystyle\int\limits_{-\infty}^{+\infty}
x\varphi (x) dx \displaystyle\frac{\psi' (0)}{1!}+\ldots .$$

Note that  $\varphi (x/\varepsilon)$ converges to the function
$$u(x)=\left\{
\begin{array}{ll}
\varphi (0), & \textrm{if} \,\, x=0,\\ 0, & \textrm{if}\,\,
x\not=0.\\
\end {array} \right.$$ Last function almost everywhere equals to zero.

Like Sobolev-Schwartz distributions we can \emph{differentiate}
$\mathbf{R}\langle\varepsilon\rangle$--distri\-bu\-tions. For
example, $$\displaystyle\int\limits_{-\infty}^{+\infty}
\frac{d}{dx}\varphi (x/\varepsilon)\psi
(x)dx=-\displaystyle\int\limits_{-\infty}^{+\infty} \varphi
(x/\varepsilon)\frac{d}{dx}\psi (x)dx,$$

$$-\displaystyle\int\limits_{-\infty}^{+\infty} \varphi
(x/\varepsilon)\frac{d}{dx}\psi (x)dx=-\varepsilon
\displaystyle\int\limits_{-\infty}^{+\infty} \varphi (x) dx \psi'
(0)-\varepsilon^2 \displaystyle\int\limits_{-\infty}^{+\infty}
x\varphi (x) dx \displaystyle\frac{\psi'' (0)}{1!}-\ldots .$$
\end{example}

It  is evident that
$\mathbf{R}\langle\varepsilon\rangle$--distributions are more
general objects than Sobo\-lev-Schwartz distributions
\cite{Schwartz}, \cite{Sobolev}.

\section{Conservation laws.\! Non-Archimedean approach.} A
conservation law asserts that the rate of change of the total
amount of substance contained in a fixed domain $G$ is equal to
the flux of that substance across the boundary of $G$. Denoting
the density of that substance by $u$, and the flux by $f$, the
conservation law is $$\frac{d}{dt}\int_{G} u(t,x) d x = -
\int_{\partial G} f \cdot \vec{n} dS.$$ Applying the divergence
theorem and taking $d/dt$ under the integral sign we obtain
$$\int_G (u_t+\mathbf{div} f) dx=0.$$ Dividing by vol $(G)$ and
shrinking $G$ to a point where all partial derivatives of $u$ and
$f$ are continuous we obtain the differential conservation law
$$u_t(t,x)+\mathbf{div} f (u(t,x))=0.$$ Note, that if $f(u)=u^2/2$
then we obtained the Hopf equation (\ref{HopfEquation}). In
general, previous calculations based on the following well known
proposition.

\begin{proposition}
If  $G\in L_{\it loc }^1(\mathbf{R})$ and
$\displaystyle\int\limits_{-\infty}^{+\infty} G(x)\psi (x)dx=0$
for any $\psi\in C_0^{\infty}(\mathbf{R})$ then $G=0$ almost
everywhere.
\end{proposition}

\begin{definition}\label{definition2}
Let us consider two sets of the smooth functions, depending on a
small parameter $\varepsilon \in (0,1]$. Let us take all functions
$v(t,x,\varepsilon)$ which have the type
$$v(t,x,\varepsilon)=l_0+\Delta l \varphi
\left(\frac{x-ct}{\varepsilon}\right),$$ $l_0, \Delta l, c$ are
real numbers, $\Delta l\not=0$ and $\varphi\in
\mathcal{S}(\mathbf{R} )$,
$\displaystyle\int\limits_{-\infty}^{+\infty} \varphi(y)dy=1$. We
denote this set of functions by $I$. We call $I$ as a set of
infinetely narrow solitons.
\end{definition}

\begin{definition}\label{definition3}
Now, let us take all functions $w(t,x,\varepsilon)$ which have the
type $$w(t,x,\varepsilon)=h_0+\Delta h H
\left(\frac{x-at}{\varepsilon}\right),$$ $h_0, \Delta h, a$ are
real numbers, $\Delta h\not=0$ and
$H(x)=\displaystyle\int\limits_{-\infty}^{x} \theta (y) d y$,
$\displaystyle\int\limits_{-\infty}^{+\infty} \theta (y) d y=1$
and $\theta\in \mathcal{S}(\mathbf{R} )$. We denote this set of
functions by $J$. We call $J$ as a set of shock waves.
\end{definition}

It is natural to consider conservation laws as an integral
expressions which contain the time $t$ as parameter. Therefore, we
introduce the following concept.

\begin{definition}\label{definition4}
The function $v\in I$ (or $w\in J$) will be a solution of the Hopf
equation up to $e^{-p}$, $p\in \mathbf{N}_0$ in the sense of
$\mathbf{R}\langle\varepsilon\rangle$--distributions if for any
$t\in [0,T]$

\begin{equation}\label{solution-1}
\int\limits_{-\infty}^{+\infty} \left\{v_t(t,x,\varepsilon )+
v(t,x,\varepsilon )v_x(t,x,\varepsilon ) \right\}\psi (x)
dx=\displaystyle\sum\limits_{k=p}^{+\infty} \xi_k \varepsilon^k\in
\mathbf{R}\langle\varepsilon\rangle,
\end{equation}

\begin{equation}\label{solution-2}
\int\limits_{-\infty}^{+\infty}\left\{w_t(t,x,\varepsilon )+
w(t,x,\varepsilon )w_x(t,x,\varepsilon ) \right\}\psi (x)
dx=\displaystyle\sum\limits_{k=p}^{+\infty} \eta_k
\varepsilon^k\in \mathbf{R}\langle\varepsilon\rangle
\end{equation}
for every $\psi\in\mathcal{S}(\mathbf{R}).$ In case when $p$ is
equal to  $+\infty$  the function $v (t,x,\varepsilon )$ (or $w
(t,x,\varepsilon )$) exactly satisfies the Hopf equation in the
sense of $\mathbf{R}\langle\varepsilon\rangle$--distributions.
\end{definition}

Certainly, one can consider instead of the Hopf equation some
conservation law.

From mathematical point of view, we deal with a infinitely
differentiable functions in definitions \ref{definition2} and
\ref{definition3}, so that we avoid the problem of distribution
multiplication. From physical point of view, functions from the
set $I$ or $J$ can describe fast processes. Mathematical models of
such processes based on functions from $I$ or $J$ may give
additional information and take in account a short zone where
physical system make a jump from one position to another.

Thus, we will consider solutions of the Hopf equation which are
infinitely narrow solitons or shock waves. It easy to see that

$$v(t,x,\varepsilon )\longrightarrow \left\{\begin{array}{ll}
l_0+\Delta l \varphi (0),& if \quad x=ct , \\ l_0 , & if \quad
x\not=ct .\end{array}\right. \quad \textrm{as} \quad
\varepsilon\to 0$$

$$w(t,x,\varepsilon )\longrightarrow h_0 +\Delta h H (x-at), \quad
\textrm{as} \quad  \varepsilon\to 0$$ $H$ is Heaviside function.

\section{Method for the numerical calculations of the
microscopic profiles of soliton like solutions of the Hopf
equation in the sense of
$\mathbf{R}\langle\varepsilon\rangle$--distributions.}

Thus, conservation laws are integral expressions. Therefore, it is
natural, that one can interpret the Hopf equation in the sense of
the definition \ref{definition4}.

We will seek a solution of the Hopf equation in the type of
infinitely narrow soliton, i.e. let us $v\in I$. Substitute
$v(t,x,\varepsilon)$ into integral expression (\ref{solution-1})
using  the following formulas
\begin{equation}
\int\limits_{-\infty}^{+\infty}\frac{\partial}{\partial
t}\left\{\varphi \left(\frac{x-c
t}{\varepsilon}\right)\right\}\psi (x)
dx=\sum\limits_{k=0}^{+\infty} c\varepsilon^{k+1} m_k
\frac{1}{k!}\psi^{(k+1)}(ct),
\end{equation}
\begin{equation}
\int\limits_{-\infty}^{+\infty}\frac{\partial}{\partial x
}\left\{\frac{1}{2}\varphi^2\left(\frac{x-c
t}{\varepsilon}\right)\right\} \psi (x) d
x=\sum\limits_{k=0}^{+\infty} -\varepsilon^{k+1} g_k
\frac{1}{k!}\psi^{(k+1)}(ct).
\end{equation}
We denote
\begin{equation}
m_k(\varphi)=\int\limits_{-\infty}^{+\infty} y^k\varphi(y) d y,
\,\,g_k(\varphi)=\int\limits_{-\infty}^{+\infty}
y^k\frac{\varphi^2(y)}{2} d y, \,\, k=0,1,2,\ldots .
\end{equation}
Thus, we obtain
\begin{equation}\label{Laurent series}
\int\limits_{-\infty}^{+\infty}\left\{ v_t +v v_x \right\}\psi  d
x=\sum\limits_{k=0}^{+\infty} \left\{\Delta l(c-l_0) m_k -(\Delta
l)^2 g_k\right\}\varepsilon^{k+1} \frac{\psi^{(k+1)}(c t)}{k!}.
\end{equation}
From the last expression we have conditions for the function
$\varphi(x)$. Namely,
\begin{equation}\label{pre-conditions}
 g_k(\varphi) - \frac{c-l_0}{\Delta l} m_k(\varphi)=0,
\,\, k=0,1,2 \ldots.
\end{equation}
From the first  ($k=0$) we have
\begin{equation}
\frac{c-l_0}{\Delta
l}=\frac{g_0}{m_0}=\frac{1}{2}\int\limits_{-\infty}^{+\infty}
\varphi^2(x) d x.
\end{equation}
Hence, we can rewrite conditions (\ref{pre-conditions}) as
follows.
\begin{equation}\label{pre-conditions-1}
\int\limits_{-\infty}^{+\infty} \varphi^2(x)d x \cdot
\int\limits_{-\infty}^{+\infty} x^k \varphi(x)d
x=\int\limits_{-\infty}^{+\infty} x^k \varphi^2(x)d x, \,\,k=0,1,2
\ldots.
\end{equation}
Now, let us prove the following lemma.
\begin{lemma}
For any non-negative integer $n$  exists such function
$\varphi\in\mathcal{S}(\mathbf{R})$, $\varphi\not\equiv 0$ which
satisfies the following system of non-linear equations:
\begin{equation}\label{conditions}
\int\limits_{-\infty}^{+\infty} x^k \varphi(x)d
x=\int\limits_{-\infty}^{+\infty} x^k \varphi^2(x)d x /
\int\limits_{-\infty}^{+\infty} \varphi^2(x)d x \,\,\,\,k=0,1,2
\ldots n.
\end{equation}
\end{lemma}
{\it Proof.} First, we will seek function  $\varphi(x)$ in the
following type:
\begin{equation}\label{function}
\varphi (x)=c_0 h_0(x)+c_1 h_1(x)+\ldots+c_{n} h_{n}(x),
\end{equation} where
\begin{equation}
h_{k}(x)=\frac{H_{k}(x)}{\sqrt{2^k k!}\sqrt[4]{\pi}} e^{-x^2/2}
\,\,\mathrm{are} \,\, \mathrm{Hermit} \,\, \mathrm{functions.}
\end{equation}
Then we substitute the expression (\ref{function}) into conditions
(\ref{conditions}). After that we will have nonlinear system of
$n+1$ equations with $n+1$ unknowns ($c_0, c_1, c_2, \ldots ,
c_n$). We write this system by the following way.

\begin{equation}\label{system}
A\vec{x}=\mathcal{N}(\vec{x}), \,\,\, \vec{x}=(c_0, c_1, \ldots ,
c_n)
\end{equation}
$A$ is a matrix with elements
$$A_{kj}=\int\limits_{-\infty}^{+\infty} x^k h_j (x) d x =
(-\mathbf{i})^j \mathbf{i}^k \sqrt{2\pi} h_j^{(k)}(0),
\,\,\,\mathbf{i}=\sqrt{-1},\,\,\, k,j=0,1,2,\ldots n $$

$\mathcal{N}$ is nonlinear map such that
\begin{equation}
\mathcal{N}(\vec{x})=\frac{1}{\|\vec{x}\|^2}\sum\limits_{k=0}^n
(N(k)\vec{x},\vec{x})\vec{e}_k\equiv \sum\limits_{k=0}^n
f_k(\vec{x})\vec{e}_k
\end{equation}
Vector $\vec{e}_k=(e_0, e_1, \ldots , e_n)$ such that $e_k=1$ and
$e_j=0$ for all $j\neq k$. $N(k)$ are matrices with elements
\begin{equation}
N_{ij}(k)=\int\limits_{-\infty}^{+\infty} x^k h_i(x) h_j(x) d x
,\,\,\, i,j,k=0,1,2 \ldots n
\end{equation}
and functions $$f_k(\vec{x})=\frac{(N(k)
\vec{x},\vec{x})}{\|\vec{x}\|^2}.$$ Note that functions
$f_k(\vec{x})$ are continuous everywhere except $\vec{x}=0$ and
$|f_k(\vec{x})|\leq \|N(k)\|$ due to Cauchy-Bunyakovskii
unequality. Matrix $A$ is invertible for any $n$ because of $\det
(A)$ is a Wronskian for the linear independent system of Hermit
functions $h_0(x)$, $h_1(x)$, ... $h_n(x)$ and $$\det
(A)=(2\pi)^\frac{(n+1)}{2} W(h_0(0), h_1(0), \ldots h_n(0)).$$ We
can write the system (\ref{system}) as
\begin{equation}\label{functionF}
\vec{x}=\sum\limits_{k=0}^n f_k(\vec{x})A^{-1}\vec{e}_k\equiv
F(\vec{x}) \,\,\,\mathrm{or}\,\,\, \vec{x}=A^{-1}
(\mathcal{N}(\vec{x}))\equiv F(\vec{x}).
\end{equation}
Let us describe the function $F:\mathbf{R}^{n+1} \longmapsto
\mathbf{R}^{n+1}$. It is continuous except $\vec{x}=0$ and
bounded. Indeed,
\begin{equation}
\|F(\vec{x})\|\leq \|A^{-1}\| \sum\limits_{k=0}^n \|N(k)\|, \,\,\,
r_n=\|A^{-1}\| \sum\limits_{k=0}^n \|N(k)\|.
\end{equation}
Let us consider function $\mathcal{N}(\vec{x})$. It is continuous
function everywhere in $\mathbf{R}^{n+1}$ except $\vec{x}=0$ and,
moreover, $\mathcal{N}(\mathbf{R}^{n+1}\backslash
\{0\})\subset\Pi_1$ where $\Pi_1=\{\vec{z}\in\mathbf{R}^{n+1}:
z_0=1 \}$ is a plane. Further $A^{-1}(\Pi_1)=\Pi_2$ where
$\Pi_2=\{\vec{y}\in\mathbf{R}^{n+1}: \sum\limits_{k=0}^n a_{0j}
y_j=1 \}$ is another plane.
$$A_{0j}=\int\limits_{-\infty}^{+\infty} h_j (x) d x =
(-\mathbf{i})^j \sqrt{2\pi} h_j(0),
\,\,\,\mathbf{i}=\sqrt{-1},\,\,\, j=0,1,2,\ldots n $$ Thus, we can
consider the function $F(\vec{x})$ which is defined on the convex
compact set $C_n=\Pi_2 \bigcap B[0,r_n]$ such that $F: C_n
\longmapsto C_n$, where $B[0,r_n]$ is a closed ball with radius
$r_n$. Function $F$ is continuous on the $C_n$ because of
$\vec{0}\not\in C_n$. Now we can use J.Schauder theorem.

\begin{theorem} [Schauder fixed-point theorem \cite{Schauder}]
Let $C$ be a compact convex subset of a normed space $E$. Then
each continuous map $F: C\longmapsto C$ has at least one fixed
point.
\end{theorem}

Hence, we can conclude that our system (\ref{functionF}) and
therefore system (\ref{system}) has at least one solution. Thus,
there is a function $\varphi(x)$ which satisfy to conditions
(\ref{conditions}) proposed lemma.

\begin{remark}
Let us a function $\varphi (x)$ satisfies lemma condition. If
$\beta\in \mathbf{R}$  then the function $\varphi (x+\beta)$ also
satisfies lemma condition. Moreover, if
$\displaystyle\int\limits_{-\infty}^{+\infty} \varphi^2 (x)\,
dx=\alpha $ then $\varphi (\alpha x)$ satisfies lemma condition.
\end{remark}

Thus, we can formulate the following result.

\begin{theorem}\label{theorem1}
For any integer $p$ there is a infinitely narrow soliton type
solution of the Hopf equation (in the sense of the definition
\ref{definition4}) up to $e^{-p}$ with respect to the norm
$|\cdot|_\nu$, i.e.
\begin{equation}
v(t,x,\varepsilon)=l_0+\Delta l \varphi
\left(\frac{x-ct}{\varepsilon}\right),
\end{equation}
$l_0, \Delta l, c$ are real numbers, $\Delta l\not=0$ and
$\varphi\in \mathcal{S}(\mathbf{R} )$,
$\displaystyle\int\limits_{-\infty}^{+\infty} \varphi(y)dy=1$.
Moreover,
\begin{equation}\label{ConstantConditions1}
\frac{c-l_0}{\Delta l}=\frac{1}{2}\int\limits_{-\infty}^{+\infty}
\varphi^2(x) d x.
\end{equation}
\end{theorem}

For example, calculations in case $p=7$ give the ``profile''
$\varphi (x)$ (see  Fig. \ref{fig1}) for the infinitely narrow
soliton $v(t,x,\varepsilon )=\varphi \left(\frac{x-c
t}{\varepsilon}\right)$:
\begin{equation}
\varphi(x)=\left\{ \frac{c_0}{\sqrt[4]{\pi}} +
\frac{c_2(4x^2-2)}{\sqrt{2^2 2!}\sqrt[4]{\pi}} +
\frac{c_4(16x^4-48x^2+12)}{\sqrt{2^4 4!}\sqrt[4]{\pi}}\right\}
e^{-x^2/2},
\end{equation}
where $c_0=0.66583$, $c_2=-0.23404$, $c_4=0.05028$, $c=0.25032$
($c$ is a velocity of the soliton). Numbers $c_0$, $c_2$, $c_4$
and $c$ were found approximately by iteration method using the
following sequence.
\begin{equation}
\vec{x}_{m+1}=A^{-1} (\mathcal{N}(\vec{x}_{m})), \,\,\,
m=0,1,2,\ldots .
\end{equation}
Matrix $A$ and a nonlinear $\mathcal{N}$ were introduced in the
lemma proof.

\begin{figure}[ht]
\begin{minipage}[b]{.49\linewidth}
\centering\includegraphics[width=\linewidth]{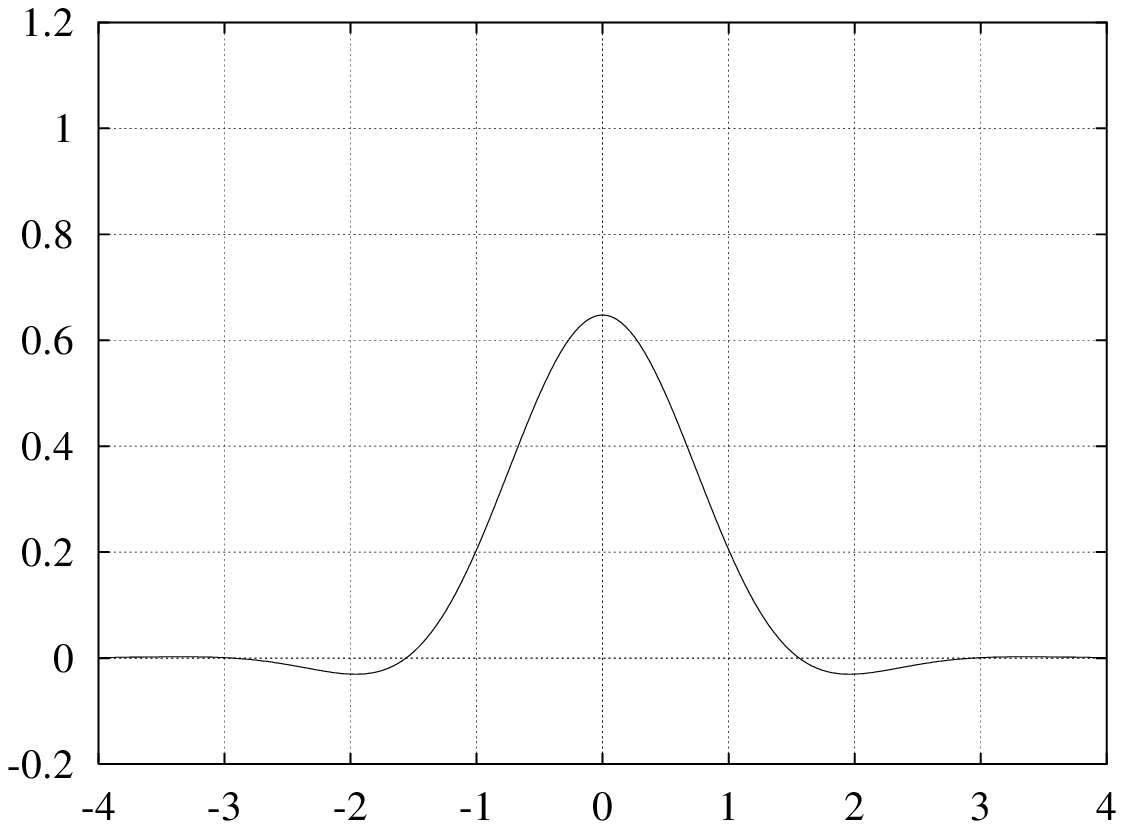}
\caption{\footnotesize The case $p=7$, $c=0.25032$.}\label{fig1}
\end{minipage}\hfill
\begin{minipage}[b]{.49\linewidth}
\centering\includegraphics[width=\linewidth]{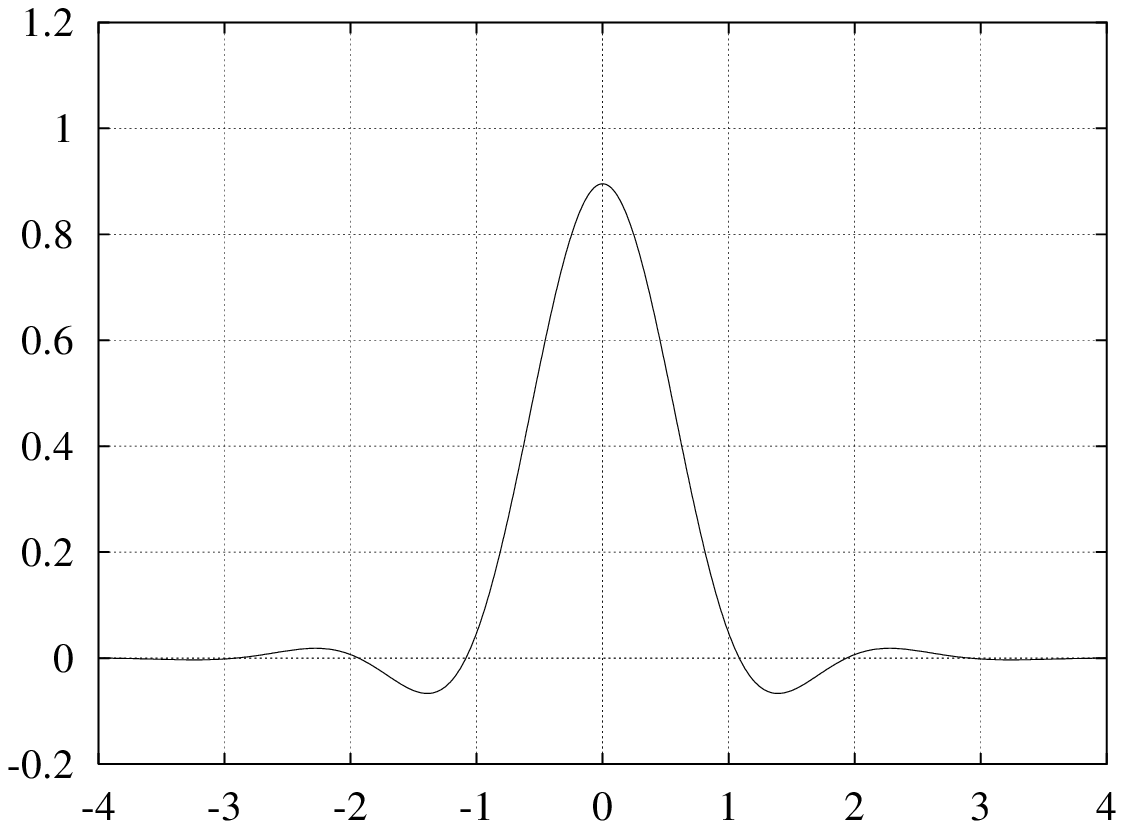}
\caption{\footnotesize The case $p=13$, $c=0.35442$.}\label{fig2}
\end{minipage}
\end{figure}

Calculations of soliton-like profiles $\varphi (x)$ for the Hopf
equation in case  $p=13,$ $15,$ $17,$ $19,$ $21$ give us pictures
(Fig. \ref{fig2}, \ref{fig3}, \ref{fig4}, \ref{fig5}, \ref{fig6}).

\begin{figure}[ht]
\begin{minipage}[b]{.49\linewidth}
\centering\includegraphics[width=\linewidth]{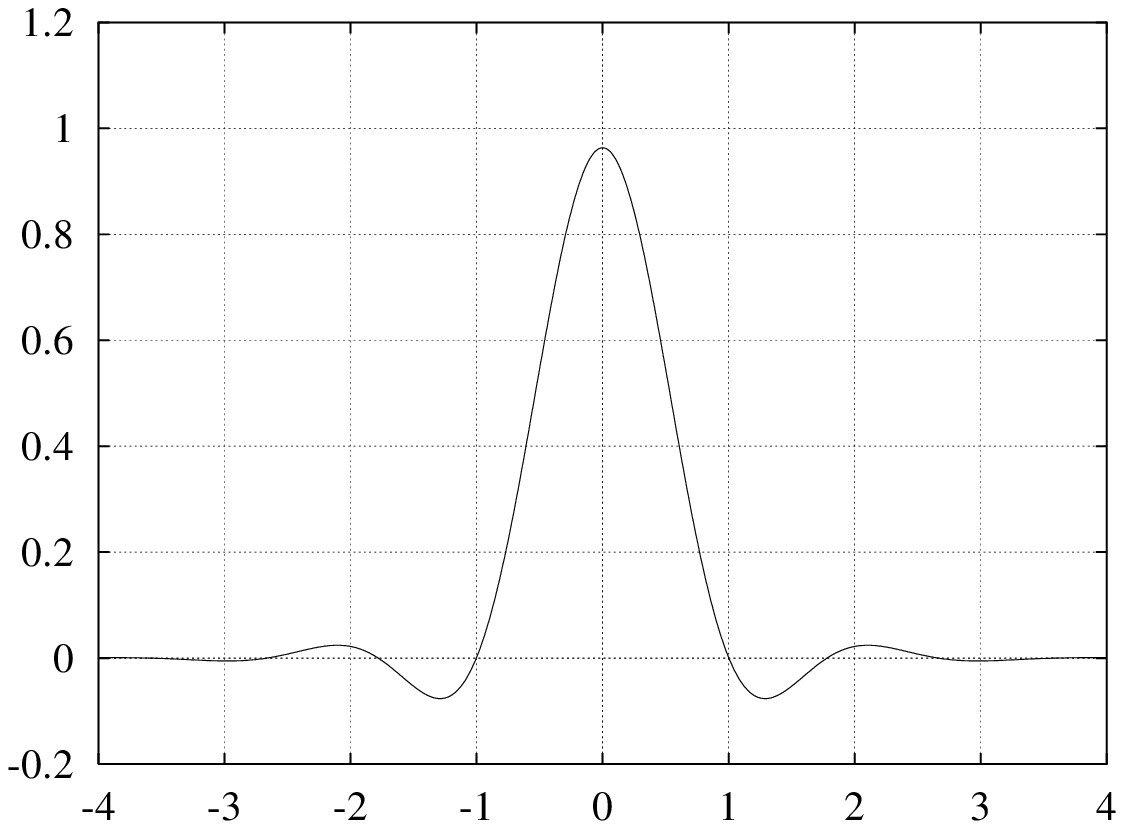}
\caption{\footnotesize The case $p=15$, $c=0.38267$.}\label{fig3}
\end{minipage}\hfill
\begin{minipage}[b]{.49\linewidth}
\centering\includegraphics[width=\linewidth]{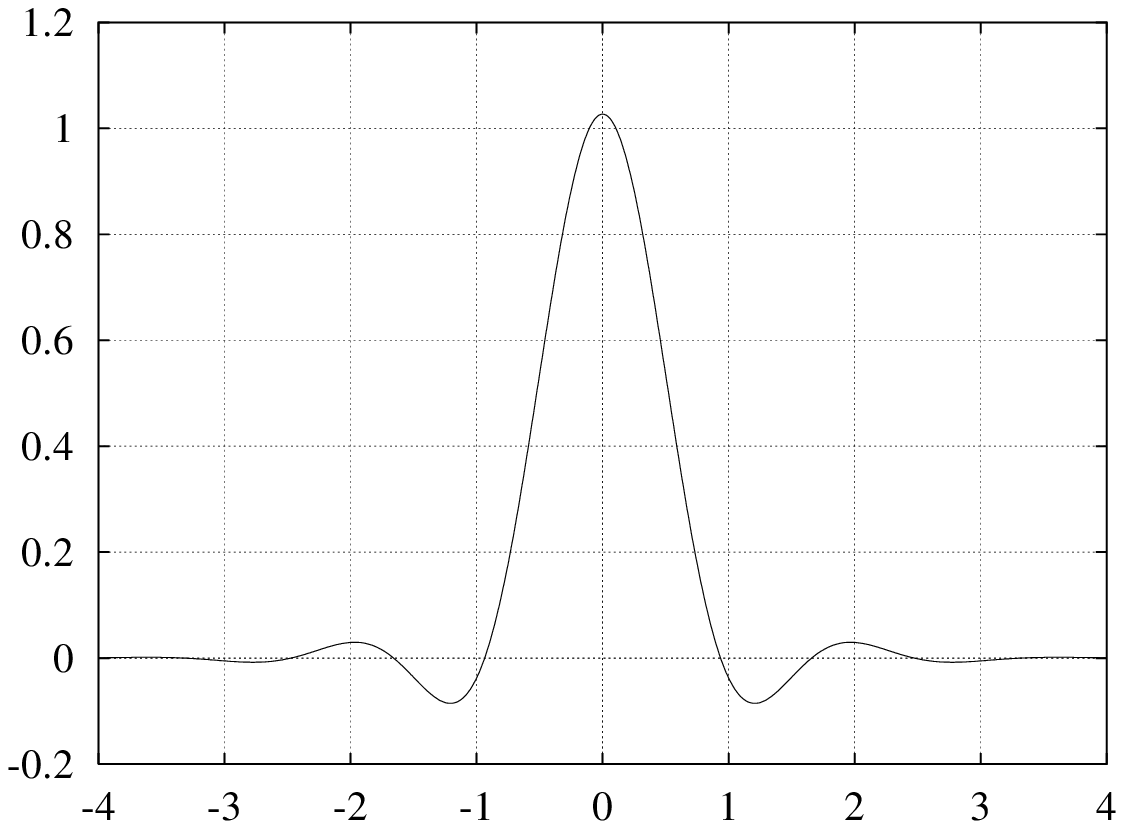}
\caption{\footnotesize The case $p=17$, $c=0.40892$.}\label{fig4}
\end{minipage}
\end{figure}

\begin{figure}[ht]
\begin{minipage}[b]{.49\linewidth}
\centering\includegraphics[width=\linewidth]{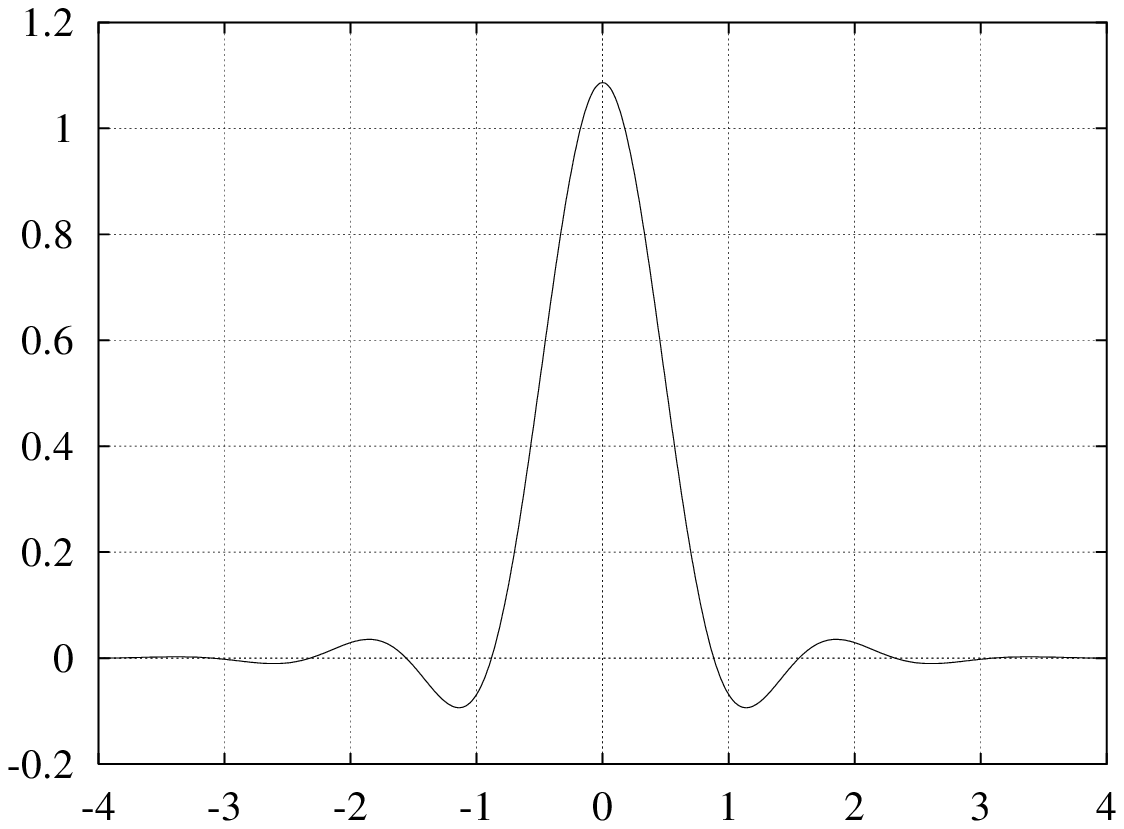}
\caption{\footnotesize The case $p=19$, $c=0.43357$.}\label{fig5}
\end{minipage}\hfill
\begin{minipage}[b]{.49\linewidth}
\centering\includegraphics[width=\linewidth]{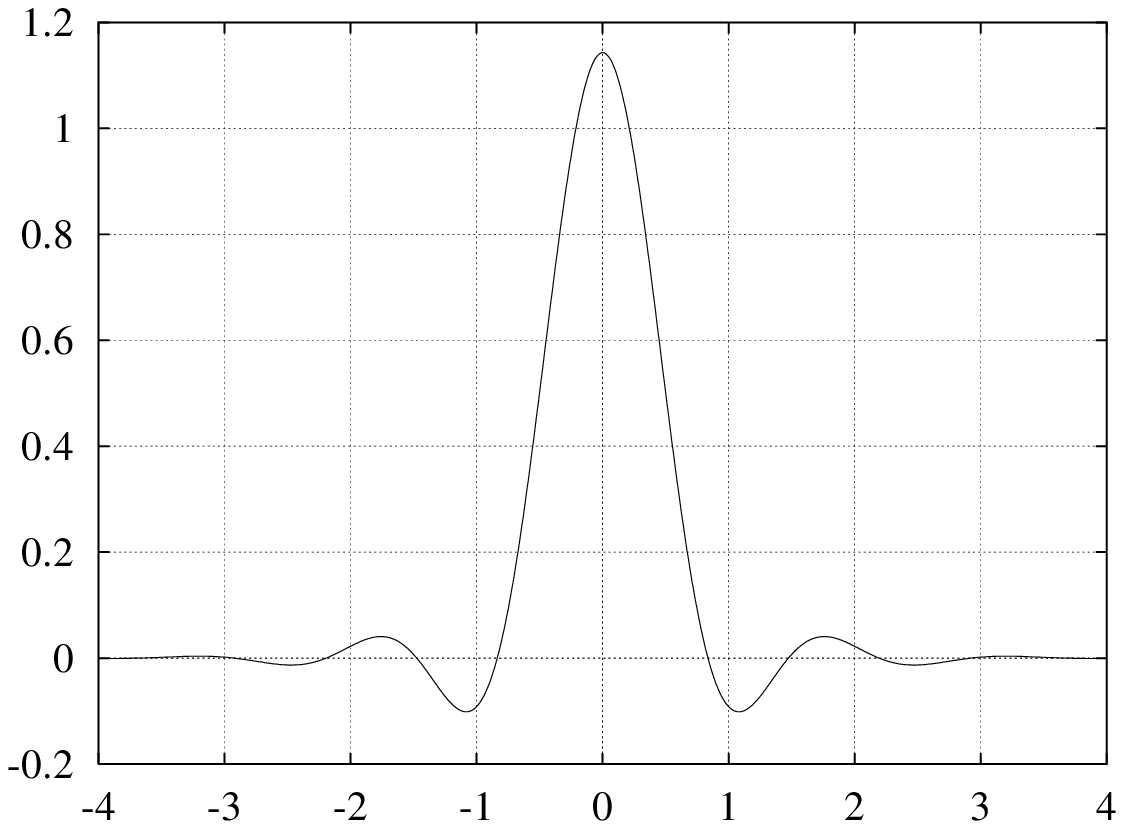}
\caption{\footnotesize The case $p=21$, $c=0.45678$.}\label{fig6}
\end{minipage}
\end{figure}

For the $p$ greater than $21$ matrix $A$ is close to singular and
calculations can be inaccurate.

\section{Calculations of the microscopic profiles of the
shock wave solutions of the Hopf equation in the sense of
$\mathbf{R}\langle\varepsilon\rangle$--distributions.}

A solution of the Hopf equation in this case we will seek in the
set $J$. Namely,

$$w(t,x,\varepsilon)=h_0+\Delta h K
\left(\frac{x-at}{\varepsilon}\right), $$ $h_0, \Delta h, a$ are
real numbers, $\Delta h\not=0$ and
$$K(x)=\int\limits_{-\infty}^{x} \theta (y) d y,
\int\limits_{-\infty}^{+\infty} \theta (y) d y=1, \quad \theta\in
\mathcal{S}(\mathbf{R} ).$$

Substitute $w(t,x,\varepsilon)$ into the integral expression
(\ref{solution-2}) using the following formulas

\begin{equation}
\int\limits_{-\infty}^{+\infty}\frac{\partial}{\partial t}\left\{K
\left(\displaystyle\frac{x-a t}{\varepsilon}\right)\right\}\psi
(x) dx=\sum\limits_{k=0}^{+\infty} (-a)\varepsilon^k m_k
\frac{\psi^{(k)}(at)}{k!},
\end{equation}
\begin{equation}
\int\limits_{-\infty}^{+\infty} K \left(\displaystyle\frac{x-a
t}{\varepsilon}\right)\frac{\partial}{\partial x}\left\{K
\left(\displaystyle\frac{x-a t}{\varepsilon}\right)\right\} \psi
(x) d x=\sum\limits_{k=0}^{+\infty} \varepsilon^k r_k
\frac{\psi^{(k)}(at)}{k!}.
\end{equation}

We denote by
\begin{equation}
m_k(\theta)=\int\limits_{-\infty}^{+\infty} y^k\theta(y) d
y,\,\,r_k(\theta)=\int\limits_{-\infty}^{+\infty} x^k \theta
(x)\left(\int\limits_{-\infty}^x \theta (y) dy\right) dx, \,\,
k=0,1,2,\ldots .
\end{equation}
Thus, we get
\begin{equation}\label{Laurent series Hopf}
\int\limits_{-\infty}^{+\infty}\left\{ w_t +ww_x\right\}\psi  d
x=\sum\limits_{k=0}^{+\infty} \left\{(\Delta h )^2 r_k-\Delta
h(a-h_0) m_k \right\}\varepsilon^{k} \frac{\psi^{(k)}(a t)}{k!}.
\end{equation}
From the last expression we have conditions for the function
$\theta(x)$
\begin{equation}\label{pre-conditions-Hopf}
r_k(\theta ) - \frac{a-h_0}{\Delta h} m_k(\theta )=0, \,\, k=0,1,2
\ldots.
\end{equation}
From the first ($k=0$) we have
\begin{equation}
\frac{a-h_0}{\Delta h}=\int\limits_{-\infty}^{+\infty} \theta (x)
\left(\int\limits_{-\infty}^x \theta (y) dy\right) dx=\frac{1}{2}.
\end{equation}
Therefore,  we can rewrite (\ref{pre-conditions-Hopf}) as
\begin{equation}\label{pre-conditions-1-Hopf}
\frac{1}{2}\int\limits_{-\infty}^{+\infty} x^k \theta(x)d
x=\int\limits_{-\infty}^{+\infty}x^k \theta (x)
\left(\int\limits_{-\infty}^x \theta (y) dy\right) dx \,\,k=0,1,2
\ldots.
\end{equation}
The same method one can prove that there is such function $\theta
(x)\in\mathcal{S}(\mathbf{R})$ which satisfies the following
conditions
\begin{equation}\label{conditions-Hopf}
\frac{1}{2}\int\limits_{-\infty}^{+\infty} x^k \theta(x)d
x=\int\limits_{-\infty}^{+\infty}x^k \theta (x)
\left(\int\limits_{-\infty}^x \theta (y) dy\right) dx \,\,k=0,1,2
\ldots n .
\end{equation}
Thus, we can formulate next result.
\begin{theorem}
For any integer $p$ there is a shock wave type solution of the
Hopf equation (in the sense of the definition \ref{definition4})
up to $e^{-p}$ with respect to the norm $|\cdot|_\nu$.
\begin{equation}
w(t,x,\varepsilon)=h_0+\Delta h K
\left(\frac{x-at}{\varepsilon}\right),
\end{equation}
$h_0, \Delta h, a$ are real numbers, $\Delta h\not=0$ and
$K(x)=\displaystyle\int\limits_{-\infty}^{x} \theta (y) d y$,
$\displaystyle\int\limits_{-\infty}^{+\infty} \theta (y) d y=1$
and $\theta\in \mathcal{S}(\mathbf{R} ).$ Moreover,
\begin{equation}\label{ConstantConditions2}
\frac{a-h_0}{\Delta h}=\frac{1}{2}.
\end{equation}
\end{theorem}

Note that the condition (\ref{ConstantConditions2}) is
\emph{Rankine --- Hugoniot condition} for the velocity of a shock
wave.

As in previous section   we  seek function  $\theta(x)$ in the
following type:
\begin{equation}
\varphi (x)=a_0 h_0(x)+a_1 h_1(x)+\ldots+a_{n} h_{n}(x),
\end{equation} where
$h_{k}(x)$ are Hermite functions. Calculations in case $p=7$ give
the following ``profile'' ($K(x)$) for the shock wave
$w(t,x,\varepsilon )=K\left(\displaystyle\frac{x-a
t}{\varepsilon}\right)$ (where $h_0=0, \Delta h=1$).

\begin{equation}\label{Firsttypeshock}
K(x)=\int\limits_{-\infty}^x \left\{ \frac{c_0}{\sqrt[4]{\pi}}
+\frac{c_2(4\tau^2-2)}{\sqrt{2^2
2!}\sqrt[4]{\pi}}+\frac{c_4(16\tau^4-48\tau^2+12)}{\sqrt{2^4
4!}\sqrt[4]{\pi}} \right\}e^{-\tau^2/2} d\tau
\end{equation}
where $c_0=0.79617$, $c_2=-0.53004$, $c_4=0.17923$, $c=1/2$ is a
velocity of the shock wave (see Fig. \ref{fig7}). Numbers $c_0$,
$c_2$, $c_4$ were found approximately.

Note that the function $K(x)$ is not unique. There is a different
function $K_1(x)$ which satisfies mentioned above conditions. It
has the following type
\begin{equation}\label{Secondtypeshock}
\begin{array}{l}
K_1(x)=\displaystyle\int\limits_{-\infty}^x
\left\{\frac{c_0}{\sqrt[4]{\pi}} +\frac{c_1 2\tau}{\sqrt{2^1
1!}\sqrt[4]{\pi}} +\frac{c_2 (4\tau^2-2)}{\sqrt{2^2
2!}\sqrt[4]{\pi}}\right\}e^{-\tau^2/2} d\tau +\\ \displaystyle
+\int\limits_{-\infty}^x\left\{\frac{c_3 (8 \tau^3-12\tau)
}{\sqrt{2^3
3!}\sqrt[4]{\pi}}+\frac{c_4(16\tau^4-48\tau^2+12)}{\sqrt{2^4
4!}\sqrt[4]{\pi}} \right\}e^{-\tau^2/2} d\tau
\end{array}
\end{equation}
where $c_0=0.18357$, $c_1=-0.73567$, $c_2=0.74733$, $c_3=0.15327$
$c_4=-0.29539$, $c=1/2$ is a velocity of the shock wave (see Fig.
\ref{fig8}). Coefficients  $c_0$, $c_1$, $c_2$, $c_3$, $c_4$ were
found approximately by the Newton iteration method.

\begin{figure}[ht]
\begin{minipage}[b]{.49\linewidth}
\centering\includegraphics[width=\linewidth]{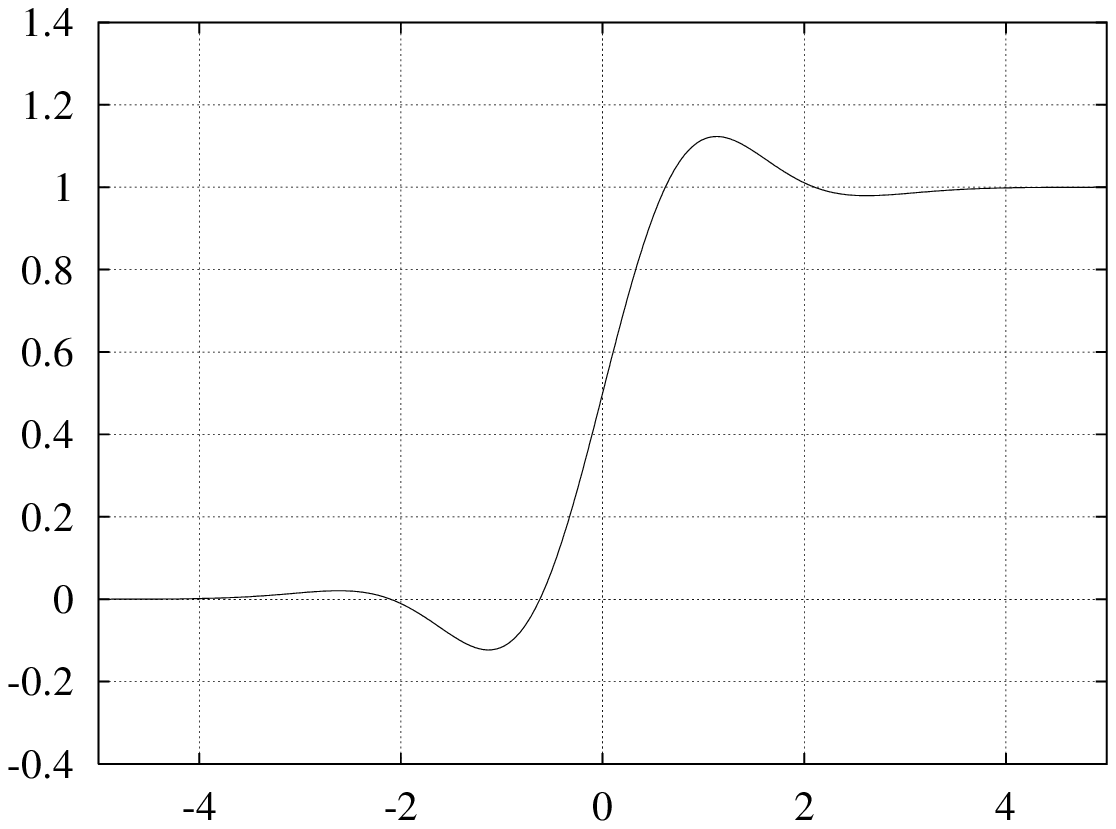}
\caption{\footnotesize Graph of the function $K(x)$.}\label{fig7}
\end{minipage}\hfill
\begin{minipage}[b]{.49\linewidth}
\centering\includegraphics[width=\linewidth]{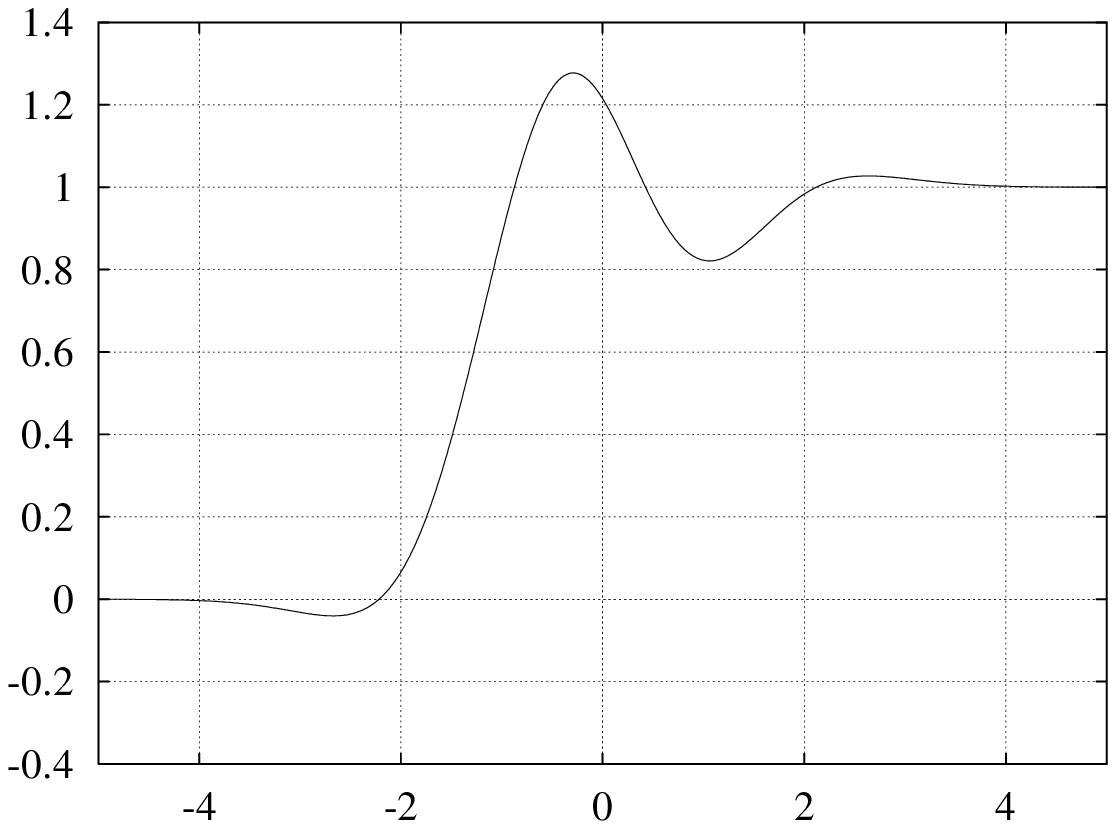}
\caption{\footnotesize Graph of the function
$K_1(x)$.}\label{fig8}
\end{minipage}
\end{figure}

Taking in account the \emph{Rankine
--- Hugoniot condition} (\ref{ConstantConditions2}) we also have
graphs (Fig. \ref{fig9}, \ref{fig10}) as a shock profiles.

\begin{figure}[ht]
\begin{minipage}[b]{.49\linewidth}
\centering\includegraphics[width=\linewidth]{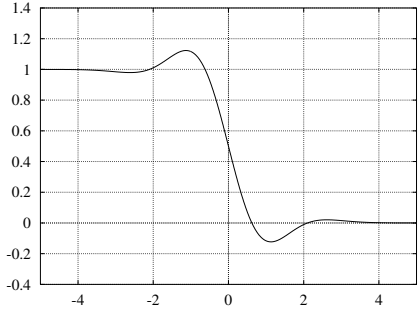}
\caption{\footnotesize First shock profile $1-K(x).$}\label{fig9}
\end{minipage}\hfill
\begin{minipage}[b]{.49\linewidth}
\centering\includegraphics[width=\linewidth]{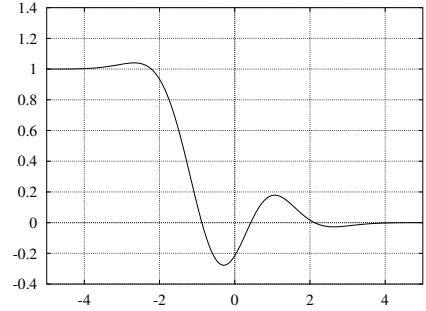}
\caption{\footnotesize Second shock profile
$1-K_1(x).$}\label{fig10}
\end{minipage}
\end{figure}

Here we describe how it is possible to find coefficients $c_0,
c_1, \ldots , c_n$  in this case by the Newton iteration method
for the following system of nonlinear equations.
\begin{equation}
P(\vec{c})=A\vec{c} -2\sum\limits_{k=0}^n
(S(k)\vec{c},\vec{c})\vec{e}_k =0, \,\,\, \vec{c}=(c_0, c_1,
\ldots , c_n)
\end{equation}
Vector $\vec{e}_k=(e_0, e_1, \ldots , e_n)$ such that $e_k=1$ and
$e_j=0$ for all $j\neq k$. $S(k)$ are matrices with elements
\begin{equation}
S_{ij}(k)=\int\limits_{-\infty}^{+\infty} x^k
h_i(x)\cdot\int\limits_{-\infty}^{x} h_j(y)\, d y\, d x,\,\,\,
i,j,k=0,1,2 \ldots n
\end{equation}

Matrix $A$ have elements
\begin{equation}
A_{ij}=\int\limits_{-\infty}^{+\infty} x^i h_j(x)\, d x ,\,\,\,
i,j=0,1,2 \ldots n
\end{equation}
We can write the formula for the Newton iteration method
\cite{Kantorovich-Akilov}.

\begin{equation}
\vec{x}_{m+1}=\vec{x}_{m}-\left[P'(\vec{x}_{m})\right]^{-1}\left[P(\vec{x}_{m})\right],
\end{equation}
where $\left[P'(\vec{x})\right]$ is a linear map depending on the
vector $\vec{x}$.
\begin{equation}
\left[P'(\vec{x})\right] [\vec{h}]=A\vec{h}
-2\left\{\sum\limits_{k=0}^n (S(k)\vec{x},\vec{h})\vec{e}_k
+\sum\limits_{k=0}^n (S^{T}(k)\vec{x},\vec{h})\vec{e}_k \right\}
\end{equation}

Calculations of shock profiles $K(x)$ for the Hopf equation in
case  $p=8,$ $9,$ $10,$ $11,$ $12,$ $13$ give us the following
pictures (Fig. \ref{fig11}, \ref{fig12}, \ref{fig13}, \ref{fig14},
\ref{fig15}, \ref{fig16}). Here, we show only two different types
of the shock type solutions of the Hopf equation. We can find more
solutions if we take a different initial data for the Newton
iteration method.

\begin{figure}[ht]
\begin{minipage}[b]{.49\linewidth}
\centering\includegraphics[width=\linewidth]{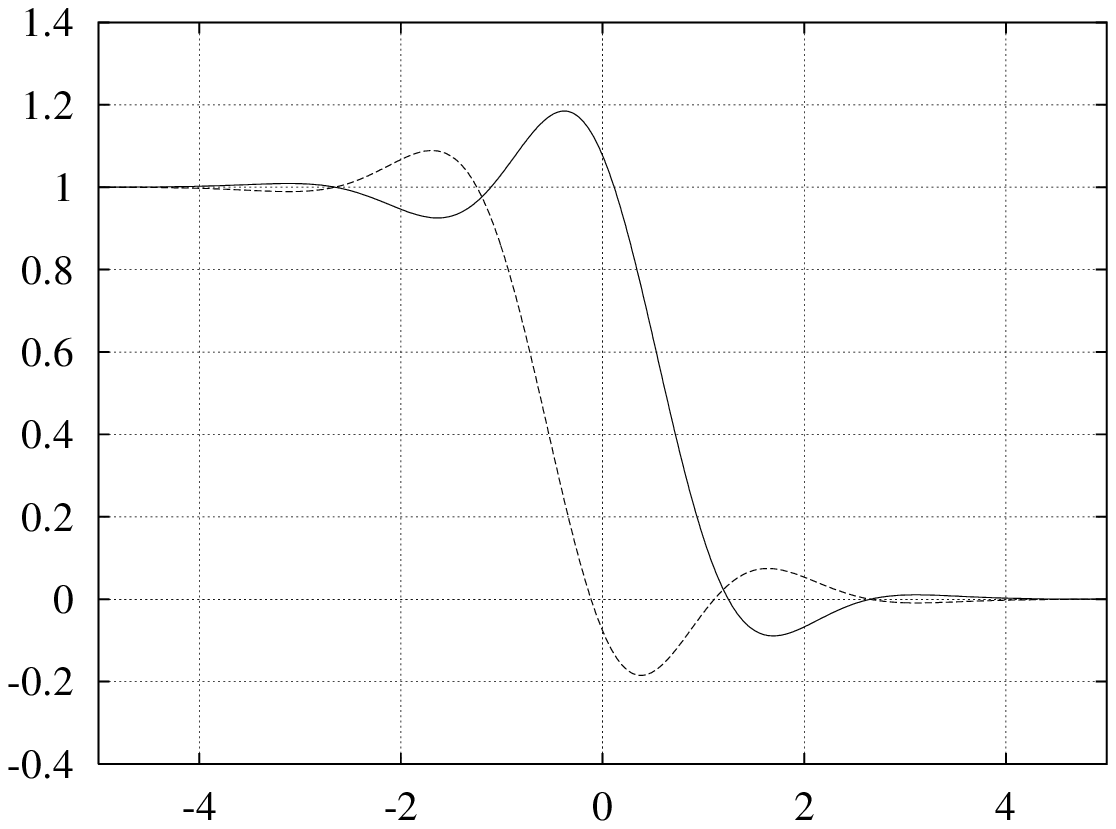}
\caption{\footnotesize Shock profiles when $p=8.$  }\label{fig11}
\end{minipage}\hfill
\begin{minipage}[b]{.49\linewidth}
\centering\includegraphics[width=\linewidth]{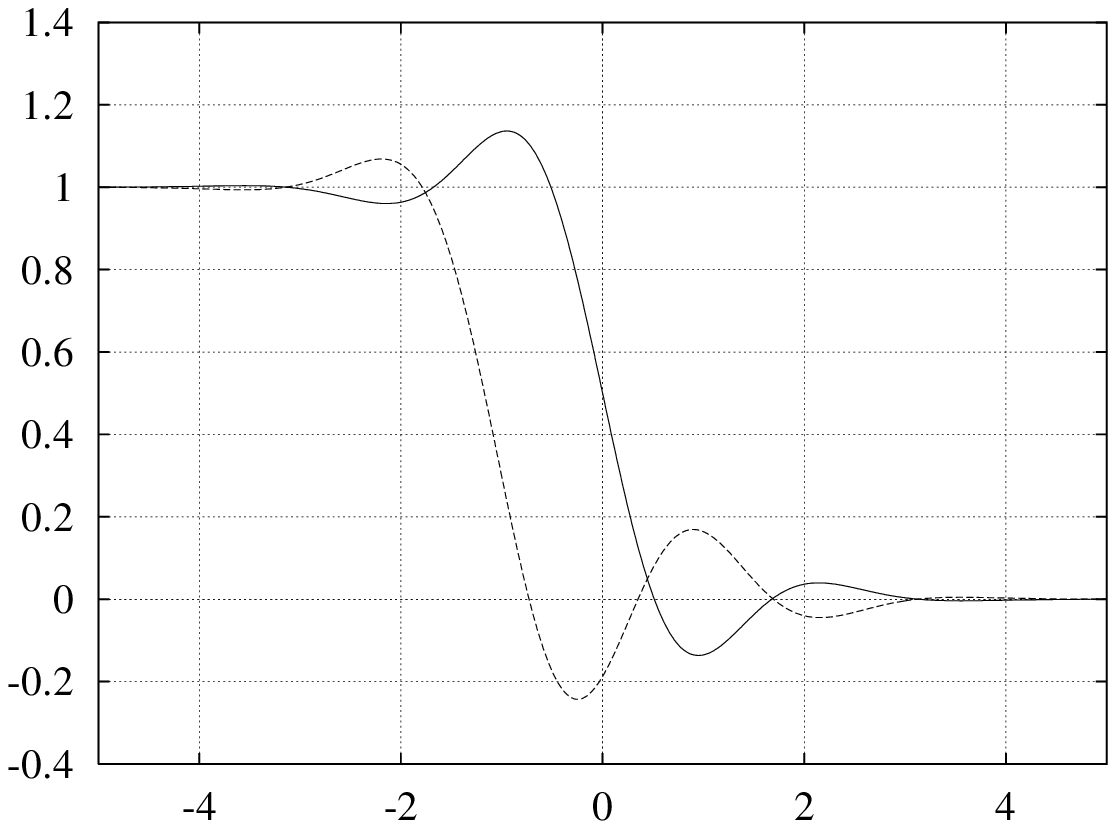}
\caption{\footnotesize Shock profiles when $p=9.$}\label{fig12}
\end{minipage}
\end{figure}

\begin{figure}[ht]
\begin{minipage}[b]{.49\linewidth}
\centering\includegraphics[width=\linewidth]{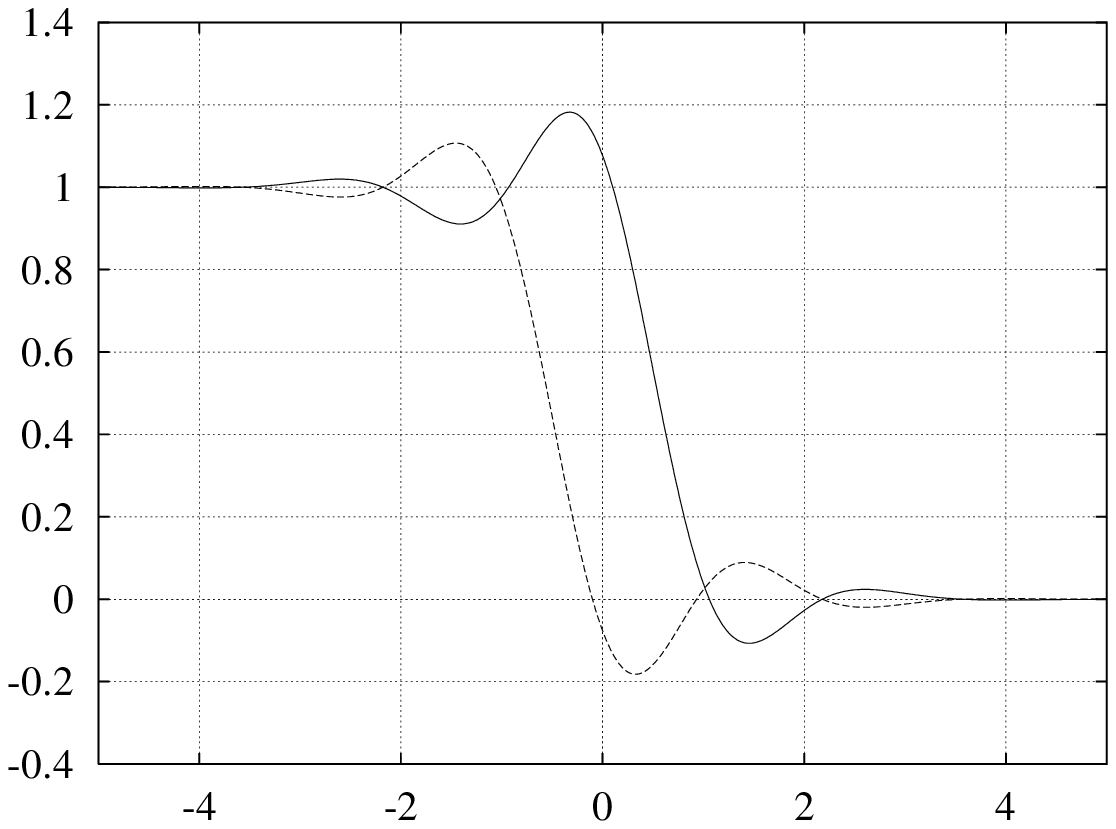}
\caption{\footnotesize Shock profiles when $p=10.$  }\label{fig13}
\end{minipage}\hfill
\begin{minipage}[b]{.49\linewidth}
\centering\includegraphics[width=\linewidth]{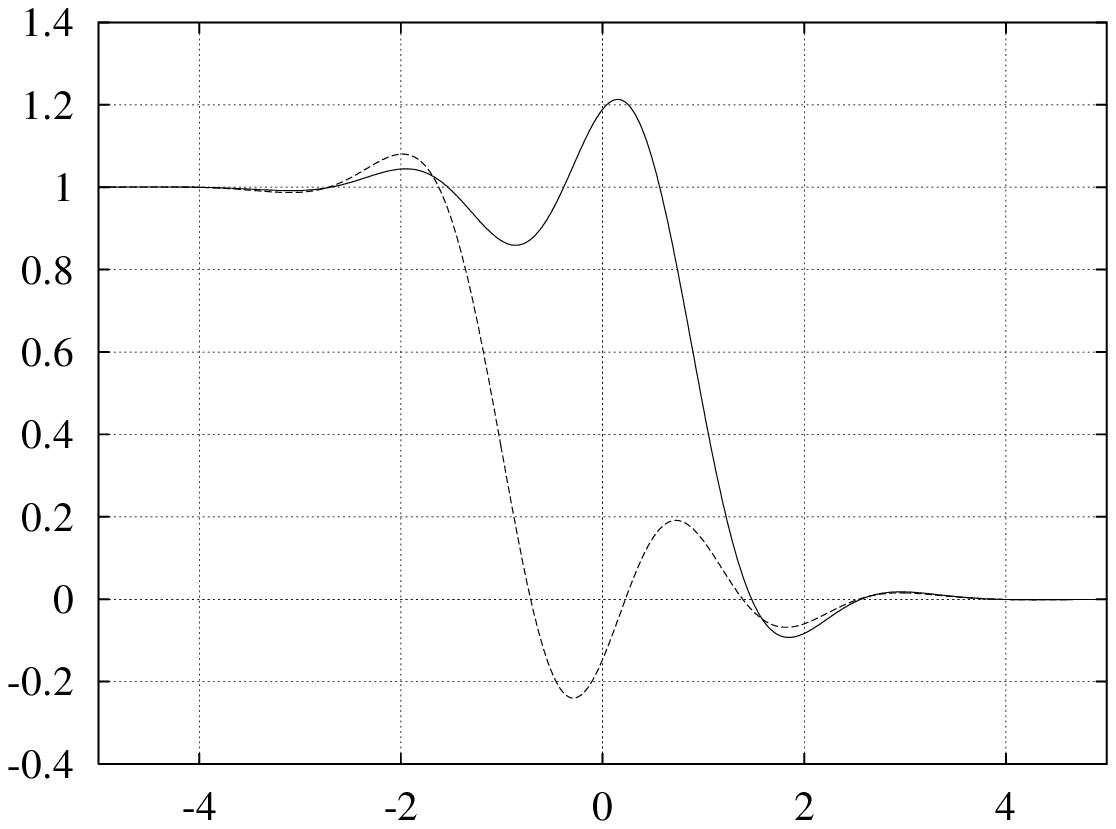}
\caption{\footnotesize Shock profiles when $p=11.$}\label{fig14}
\end{minipage}
\end{figure}

\begin{figure}[ht]
\begin{minipage}[b]{.49\linewidth}
\centering\includegraphics[width=\linewidth]{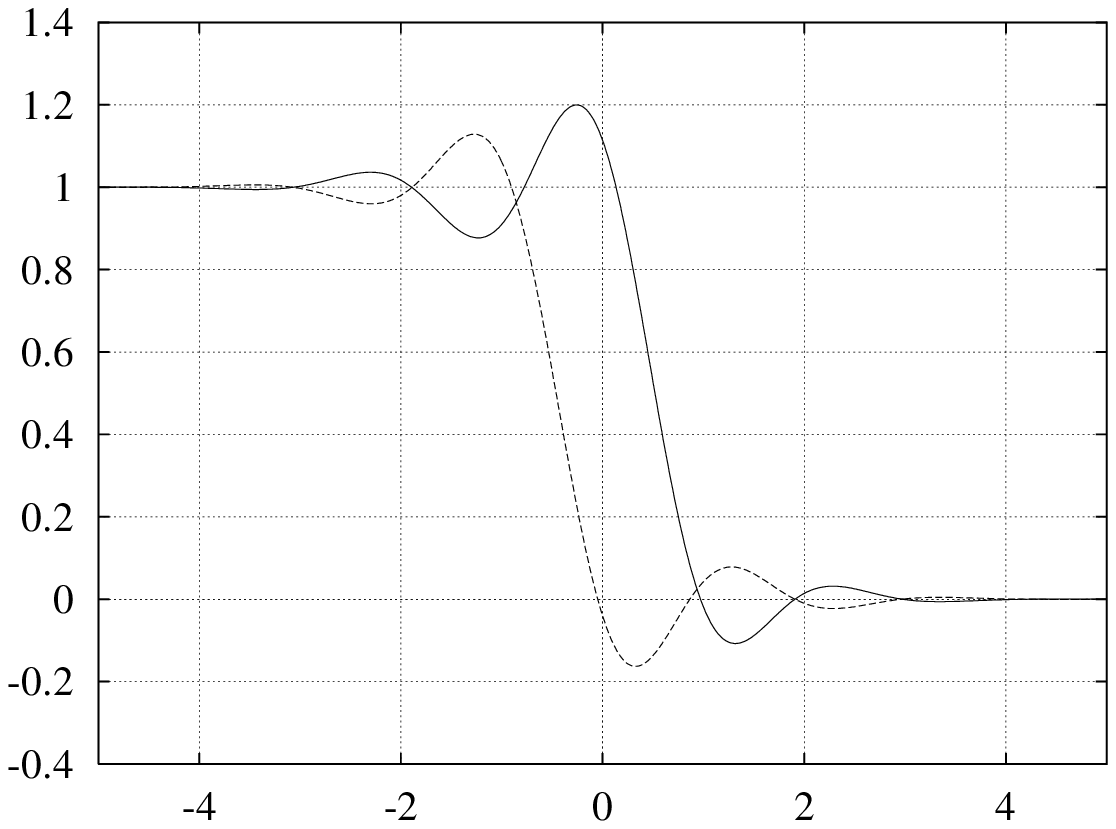}
\caption{\footnotesize Shock profiles when $p=12.$ }\label{fig15}
\end{minipage}\hfill
\begin{minipage}[b]{.49\linewidth}
\centering\includegraphics[width=\linewidth]{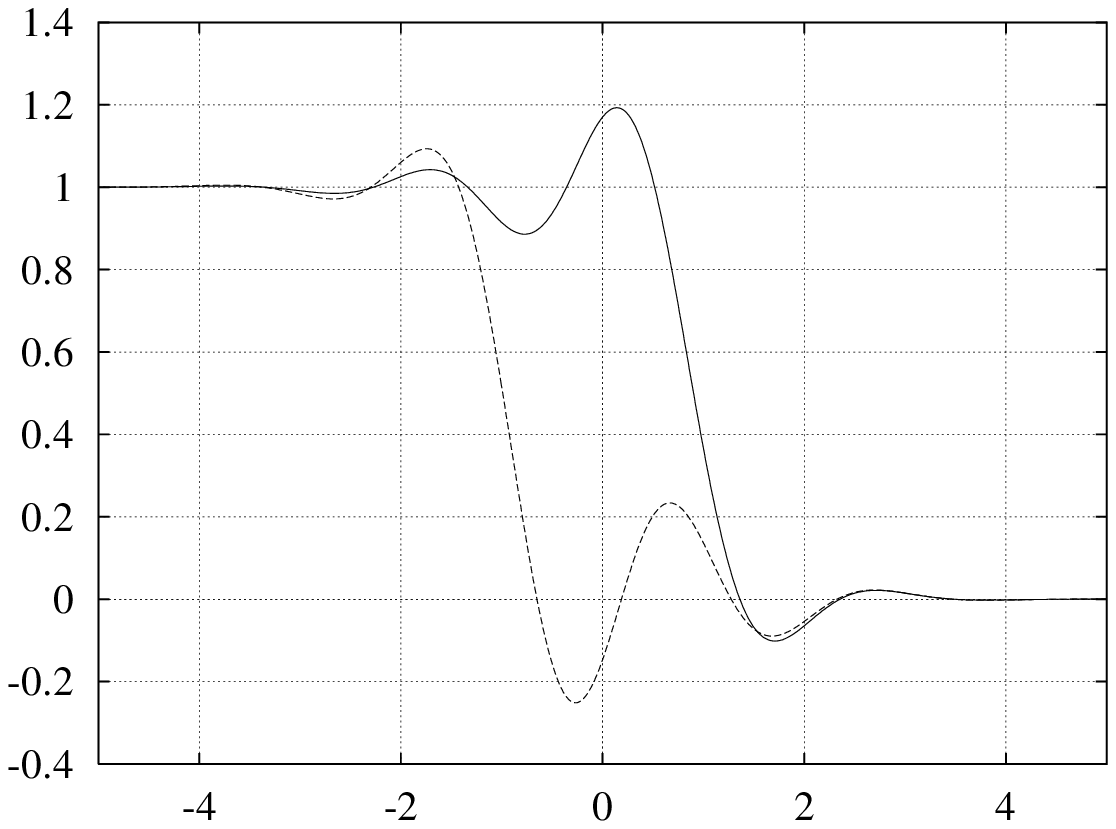}
\caption{\footnotesize Shock profiles when $p=13.$}\label{fig16}
\end{minipage}
\end{figure}

\begin{remark}
It is not easy to see  that there is exist function
\begin{equation}
\theta(x)=\sum\limits_{n=1}^{\infty} a_n h_n(x),\,\,\,
\vec{a}=(a_0, a_1, \ldots , a_n ,\ldots)\in l_2,
\end{equation}
such that
\begin{equation}
\frac{1}{2}\int\limits_{-\infty}^{+\infty} x^k \theta(x)d
x=\int\limits_{-\infty}^{+\infty}x^k \theta (x)
\left(\int\limits_{-\infty}^x \theta (y) dy\right) dx, \,\,k=0,1,2
\ldots.
\end{equation}
We think that it is true.
\end{remark}


\section{Calculations of the microscopic profiles of the shock wave
solutions of equations of elasticity theory in the sense of
$\mathbf{R}\langle\varepsilon\rangle$--distributions.} Let us
consider the following  system.
\begin{equation}\label{system-1}
\begin{array}{l}
u_t+ (u^2)_x=\sigma_x \,\,\,(\textrm{the conservation law for
momentum})
\\ \sigma_t+ u\sigma_x=k^2 u_x \,\,\, (\textrm{the Hooke law})
\end{array}
\end{equation}
Here, $u$ is the velocity of a medium and $\sigma$ is the stress.
We suppose that density of a medium is equal to the constant $1$
and $k^2$ some constant.

We will seek for a solution of this system in the following form
\begin{equation}\label{velocity-function}
u(t,x,\varepsilon)=u_0+\Delta u U
\left(\frac{x-vt}{\varepsilon}\right),
\end{equation}
$u_0, \Delta u, v$ are real numbers, $\Delta u\not=0$ and
$U(x)=\displaystyle\int\limits_{-\infty}^{x} \widetilde{U} (y) d
y$, $\displaystyle\int\limits_{-\infty}^{+\infty} \widetilde{U}
(y) d y=1$ and $\widetilde{U}\in \mathcal{S}(\mathbf{R} ).$

\begin{equation}\label{stress-function}
\sigma(t,x,\varepsilon)=\sigma_0+\Delta \sigma \Sigma
\left(\frac{x-vt}{\varepsilon}\right),
\end{equation}
$\sigma_0, \Delta \sigma, v$ are real numbers, $\Delta
\sigma\not=0$ and
$\Sigma(x)=\displaystyle\int\limits_{-\infty}^{x}
\widetilde{\Sigma} (y) d y$,
$\displaystyle\int\limits_{-\infty}^{+\infty} \widetilde{\Sigma}
(y) d y=1$ and $\widetilde{\Sigma}\in \mathcal{S}(\mathbf{R} ).$
Note that $v$ is a velocity of the shock waves.

In the other hand, we suppose
\begin{equation}\label{velocity-expansion}
\widetilde{U}(x)=a_0 h_0(x)+a_1 h_1(x)+\ldots+a_{n}
h_{n}(x),\,\,\, \vec{a}=(a_0, a_1, \ldots , a_n),
\end{equation}
\begin{equation}\label{stress-expansion}
\widetilde{\Sigma}(x)=c_0 h_0(x)+c_1 h_1(x)+\ldots+c_{n} h_{n}(x),
\,\,\,\vec{c}=(c_0, c_1, \ldots , c_n)
\end{equation}
where $h_{k}(x)$ are Hermite functions.

We understand the  solution of the system in sense of
$\mathbf{R}\langle\varepsilon\rangle$--distributions.

\begin{definition}\label{definition5}
Functions $u\in J$ and $\sigma\in J$ is a solution of the system
(\ref{system-1}) up to $e^{-p}$, $p\in \mathbf{N}_0$ in the sense
of $\mathbf{R}\langle\varepsilon\rangle$--distributions if for any
$t\in [0,T]$
\end{definition}

\begin{equation}\label{solution-1ofthesystem-1}
\int\limits_{-\infty}^{+\infty} \left\{u_t(t,x,\varepsilon )+
2u(t,x,\varepsilon ) u_x(t,x,\varepsilon
)-\sigma_x(t,x,\varepsilon )\right\}\psi (x)
dx=\displaystyle\sum\limits_{k=p}^{+\infty} \xi_k \varepsilon^k\in
\mathbf{R}\langle\varepsilon\rangle,
\end{equation}

\begin{equation}\label{solution-2ofthesystem-1}
\int\limits_{-\infty}^{+\infty}\left\{\sigma_t(t,x,\varepsilon )+
u(t,x,\varepsilon )\sigma_x(t,x,\varepsilon )-k^2
u_x(t,x,\varepsilon )\right\}\psi (x)
dx=\displaystyle\sum\limits_{k=p}^{+\infty} \eta_k
\varepsilon^k\in \mathbf{R}\langle\varepsilon\rangle
\end{equation}
for every $\psi\in\mathcal{S}(\mathbf{R} )$.

In case when $p$ is equal to  $+\infty$  functions $u
(t,x,\varepsilon )$ and   $\sigma (t,x,\varepsilon )$) exactly
satisfies the system (\ref{system-1}) in the sense of
$\mathbf{R}\langle\varepsilon\rangle$--distributions.

Substituting $u$ and $\sigma$ into
(\ref{solution-1ofthesystem-1}), (\ref{solution-2ofthesystem-1})
we get the following relations for the moments.

\begin{equation}\label{system-1conditions-1}
\{2u_0 \Delta u -v\Delta u\} m_{k}(\widetilde{U}) + 2(\Delta u)^2
m_k (\widetilde{U}U)-\Delta \sigma m_k(\widetilde{\Sigma})=0,
\,\,\,k=0,1,2,\ldots n
\end{equation}
\begin{equation}\label{system-1conditions-2}
\{u_0 \Delta\sigma  -v\Delta\sigma \} m_{k}(\widetilde{\Sigma})+
\Delta u\Delta\sigma m_k (\widetilde{\Sigma} U)-k^2 \Delta u
m_k(\widetilde{U})=0, \,\,\,k=0,1,2,\ldots n
\end{equation}

We denote as usual by
\begin{equation}
m_k(\widetilde{U})=\int\limits_{-\infty}^{+\infty} x^k
\widetilde{U}(x) dx,
\end{equation}

\begin{equation}
m_k(\widetilde{U}U)=\int\limits_{-\infty}^{+\infty} x^k
\widetilde{U} (x)\left(\int\limits_{-\infty}^x \widetilde{U} (y)
dy\right) dx, \,\, k=0,1,2,\ldots ,
\end{equation}

\begin{equation}
m_k(\widetilde{\Sigma}U)=\int\limits_{-\infty}^{+\infty} x^k
\widetilde{\Sigma} (x)\left(\int\limits_{-\infty}^x \widetilde{U}
(y) dy\right) dx, \,\, k=0,1,2,\ldots
\end{equation}

It is easy to find $v$ from (\ref{system-1conditions-1}) when k=0.
Indeed, $$\{2u_0 \Delta u -v\Delta u\}  + (\Delta u)^2 -\Delta
\sigma =0.$$ Therefore,
\begin{equation}\label{shock-velocity1}
v=2u_0+\Delta u -\frac{\Delta \sigma}{\Delta u}.
\end{equation}
Substitute $v$ into the (\ref{system-1conditions-1}). We have
\begin{equation}\label{system-1conditions-1'}
\{\Delta \sigma -(\Delta u)^2\} m_{k}(\widetilde{U}) + 2(\Delta
u)^2 m_k (\widetilde{U}U)-\Delta \sigma m_k(\widetilde{\Sigma})=0,
\,\,\,k=0,1,2,\ldots n
\end{equation}
Because of $\Delta U$ and $\Delta\sigma$ some real numbers,
therefore, all three vectors with coordinates
$m_{k}(\widetilde{U})$, $m_{k}(\widetilde{U}U)$ and
$m_k(\widetilde{\Sigma})$, $k=0,1,2,\ldots n$, respectively should
be collinear. However,
$$m_{0}(\widetilde{U})=m_{0}(\widetilde{\Sigma})=1.$$ Hence,

$$m_{k}(\widetilde{U})=m_{k}(\widetilde{\Sigma}), \,\,\,
k=0,1,2,\ldots n.$$

Thus, $\vec{a}=\vec{c}$ and from (\ref{system-1conditions-1'})
follows that $$m_{k}(\widetilde{U})=2m_{k}(\widetilde{U}U), \,\,\,
k=0,1,2,\ldots n.$$ This system we already know how to solve by
the Newton iteration method. See conditions
(\ref{conditions-Hopf}) and solution in this case.

Substitute $v$ into the (\ref{system-1conditions-2})and take into
account previous equalities we have
\begin{equation}\label{system-1conditions-2'}
\left\{\frac{(\Delta\sigma)^2}{\Delta u}-u_0 \Delta\sigma -\Delta
u \Delta\sigma \right\} m_{k}(\widetilde{\Sigma})+ \Delta
u\Delta\sigma m_k (\widetilde{\Sigma} \Sigma)-k^2 \Delta u
m_k(\widetilde{\Sigma})=0
\end{equation}
where $k=0,1,2,\ldots n$. The last expression gives us relation
for constants $\Delta\sigma$, $\Delta u$, $u_0$, $k^2$. Namely,
\begin{equation}
\left\{\frac{(\Delta\sigma)^2}{\Delta u}-u_0 \Delta\sigma -\Delta
u \Delta\sigma \right\} +\frac{1}{2}\Delta u\Delta\sigma -k^2
\Delta u =0
\end{equation}
or
\begin{equation}
(\Delta\sigma)^2 -\left(u_0+\frac{1}{2}\Delta u \right)\Delta u
\Delta\sigma -k^2 (\Delta u)^2 =0
\end{equation}

If $\Delta u$, $u_0$, $k^2$ are known then from the last equation
one can find $\Delta \sigma$

\begin{equation}
\Delta\sigma_{1,2}=\frac{1}{2} \left(u_0+\frac{1}{2} \Delta u
\right)\pm\frac{1}{2}|\Delta u| \sqrt{(u_0+\frac{1}{2}\Delta u
)^2+4k^2}
\end{equation}

In particular, if $\Delta u=-1$, $u_0=1$ then
$$\Delta\sigma_{1,2}=-\frac{1}{4}\pm\frac{1}{4}\sqrt{1+16k^2}.$$

Shock profiles of the considered system (\ref{system-1}) one can
find on pictures (Fig. \ref{fig17}, \ref{fig18}, \ref{fig19},
\ref{fig20}). We considered case when $p=13$ and $\Delta u=-1$,
$u_0=1$, $k^2=0.1$. We can also take any real $\sigma_0$ but here
we took $\sigma_0=0.5$ and then calculated $\Delta\sigma_1$ (Fig.
\ref{fig17}, \ref{fig19}, the velocity of shocks is $v=1.1531$ )
and $\Delta\sigma_2$ (Fig. \ref{fig18}, \ref{fig20}, the velocity
of shocks is $v=0.34689$). We consider two different types of
shock profiles. The first is on the Fig. \ref{fig17}, \ref{fig18}.
The second is on the Fig. \ref{fig19}, \ref{fig20}.

\begin{theorem}
For any integer $p$ there is a solution of the system of equations
(\ref{system-1}) in the sense of the definition \ref{definition5}.
Moreover, $$v=2u_0+\Delta u -\frac{\Delta \sigma}{\Delta u} $$ and
$$(\Delta\sigma)^2 -\left(u_0+\frac{1}{2}\Delta u \right)\Delta u
\Delta\sigma -k^2 (\Delta u)^2 =0 .$$
\end{theorem}

\begin{figure}[ht]
\begin{minipage}[b]{.49\linewidth}
\centering\includegraphics[width=\linewidth]{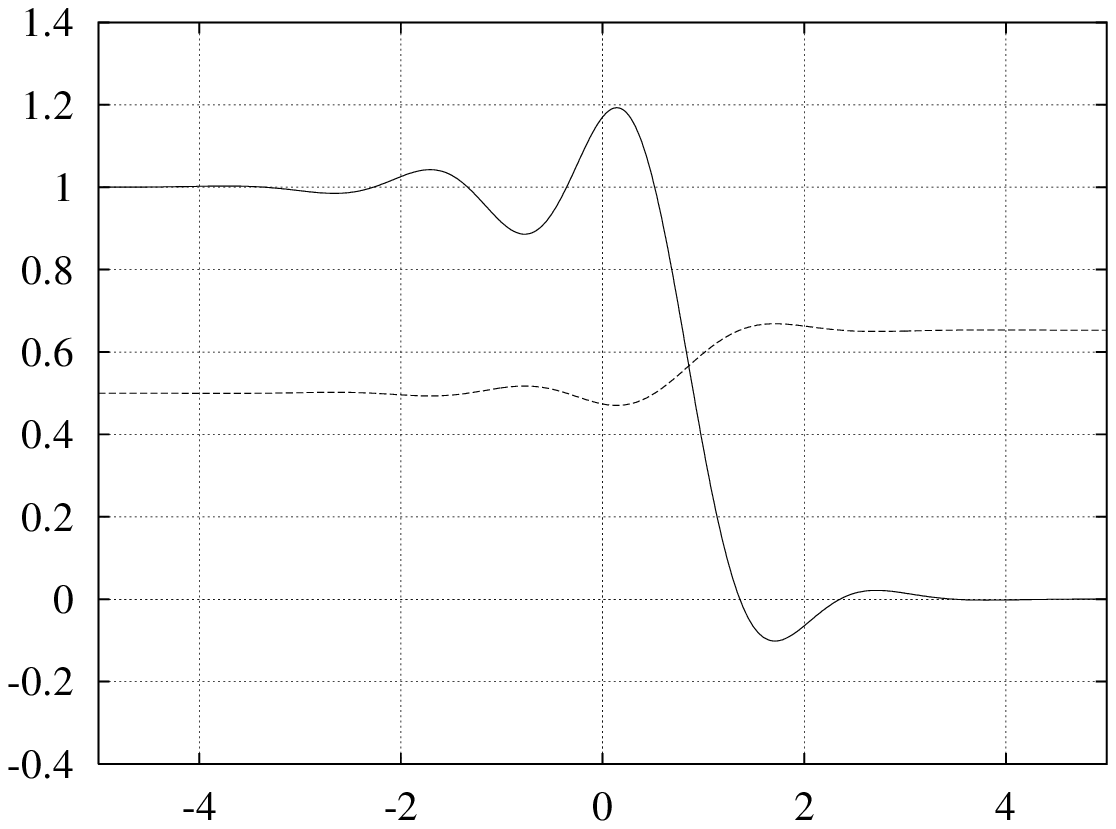}
\caption{\footnotesize Shock profiles of velocity and
stress.}\label{fig17}
\end{minipage}\hfill
\begin{minipage}[b]{.49\linewidth}
\centering\includegraphics[width=\linewidth]{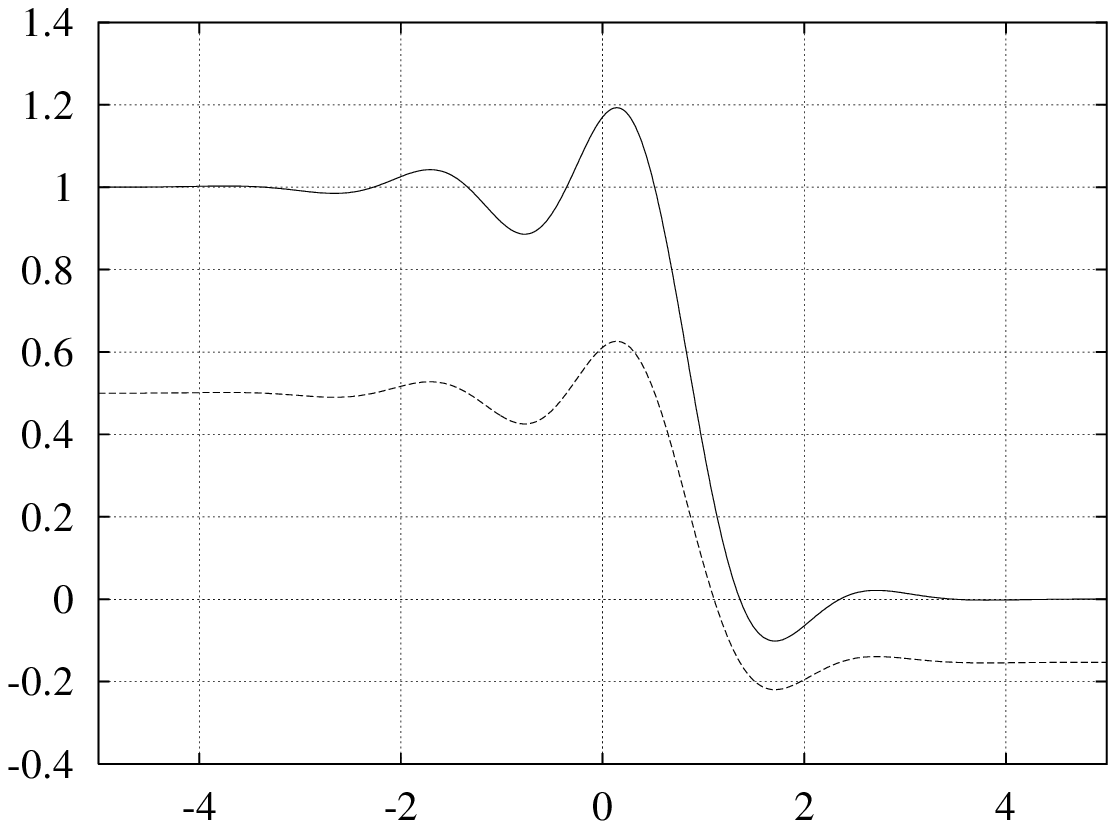}
\caption{\footnotesize Shock profiles of velocity and
stress.}\label{fig18}
\end{minipage}
\end{figure}

\begin{figure}[ht]
\begin{minipage}[b]{.49\linewidth}
\centering\includegraphics[width=\linewidth]{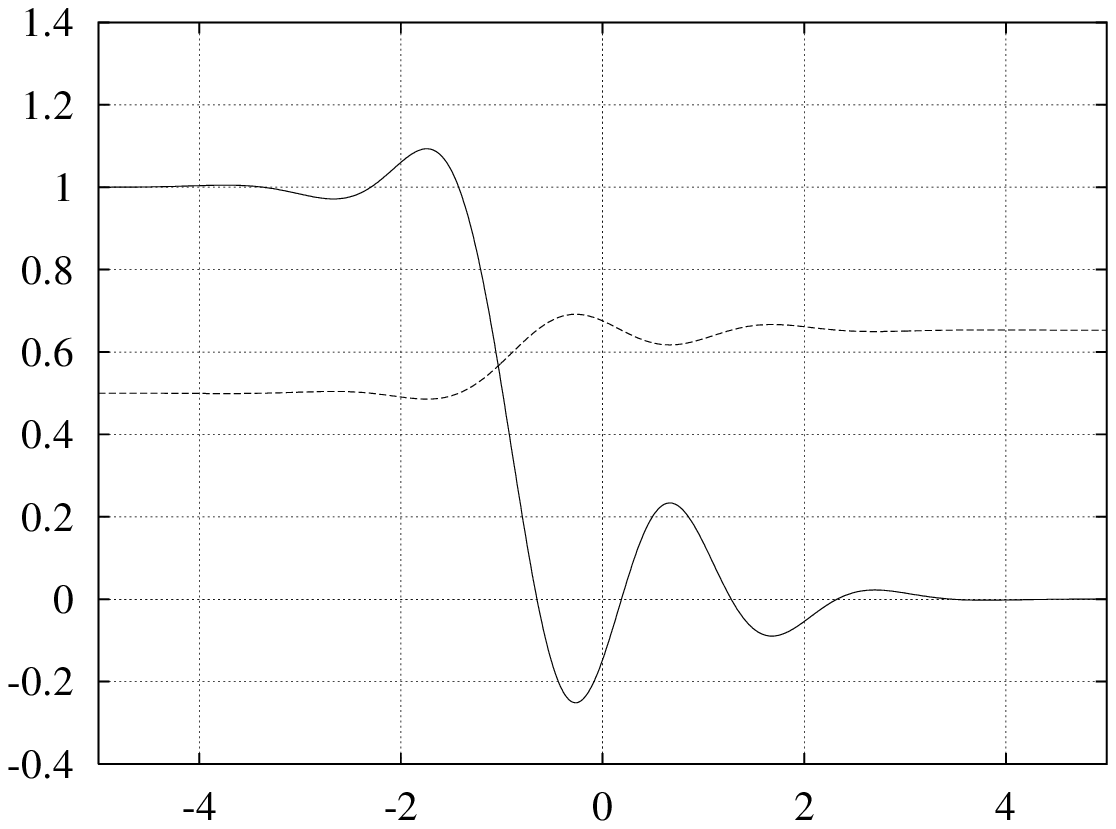}
\caption{\footnotesize Shock profiles of velocity and
stress.}\label{fig19}
\end{minipage}\hfill
\begin{minipage}[b]{.49\linewidth}
\centering\includegraphics[width=\linewidth]{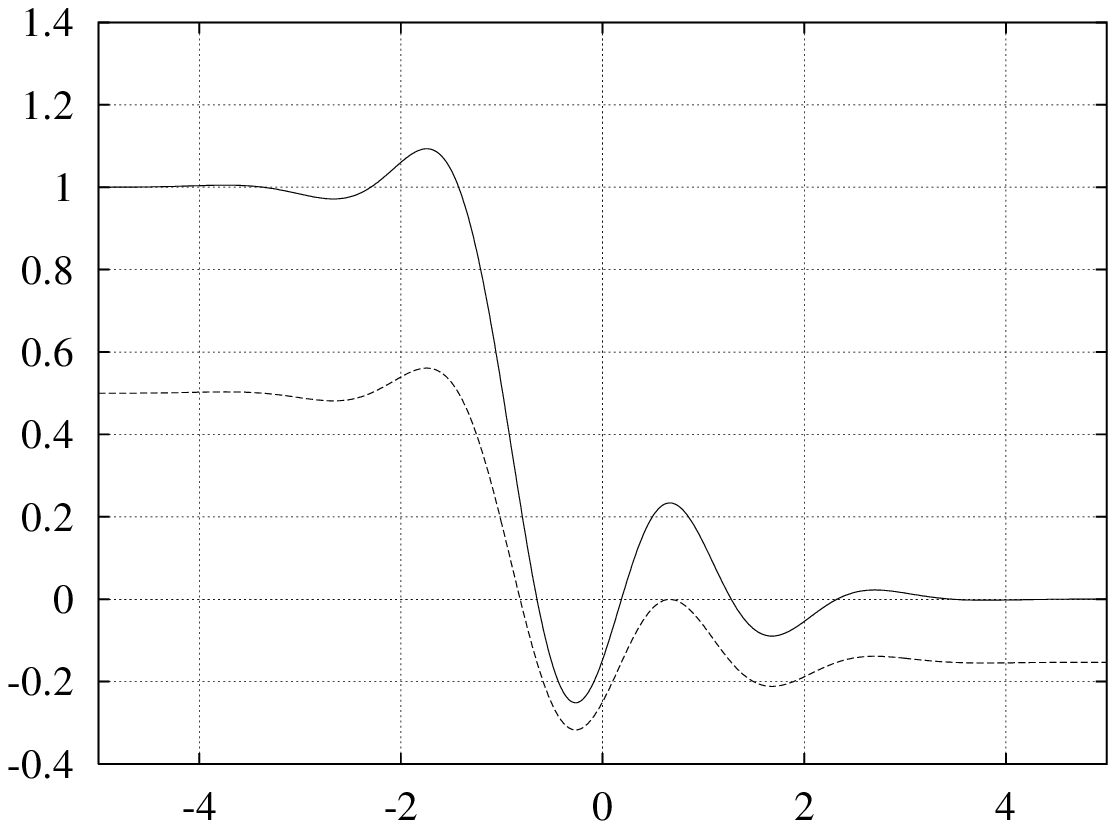}
\caption{\footnotesize Shock profiles of velocity and
stress.}\label{fig20}
\end{minipage}
\end{figure}

Let us consider the following  system.
\begin{equation}\label{system-2}
\begin{array}{l}
\rho_t+(\rho u)_x=0\,\,\,(\textrm{the conservation law for
mass})\\
 (\rho u)_t+ (\rho u^2)_x=\sigma_x \,\,\,(\textrm{the
conservation law for momentum})
\\ \sigma_t+ u\sigma_x=k^2 u_x \,\,\, (\textrm{the Hooke law})
\end{array}
\end{equation}
Here, $u$ is the velocity of a medium and $\sigma$ is the stress.
We suppose that  $k^2$ is some constant.

\begin{definition}\label{definition6}
Functions $u\in J$, $\rho\in J$ and $\sigma\in J$ is a solution of
the system (\ref{system-2}) up to $e^{-p}$, $p\in \mathbf{N}_0$ in
the sense of $\mathbf{R}\langle\varepsilon\rangle$--distributions
if for any $t\in [0,T]$

\begin{equation}\label{solution-1ofthesystem-2}
\int\limits_{-\infty}^{+\infty} \left\{\rho_t(t,x,\varepsilon )+
\rho_x
  u  +\rho
u_x \right\}\psi (x) dx=\displaystyle\sum\limits_{k=p}^{+\infty}
\xi_k \varepsilon^k\in \mathbf{R}\langle\varepsilon\rangle,
\end{equation}

\begin{equation}\label{solution-2ofthesystem-2}
\int\limits_{-\infty}^{+\infty} \left\{\rho_t u +\rho u_t +\rho_x
u^2
 +2\rho  u u_x -\sigma_x \right\}\psi (x)
dx=\displaystyle\sum\limits_{k=p}^{+\infty} \zeta_k
\varepsilon^k\in \mathbf{R}\langle\varepsilon\rangle,
\end{equation}

\begin{equation}\label{solution-3ofthesystem-2}
\int\limits_{-\infty}^{+\infty}\left\{\sigma_t + u \sigma_x -k^2
u_x \right\}\psi (x) dx=\displaystyle\sum\limits_{k=p}^{+\infty}
\eta_k \varepsilon^k\in \mathbf{R}\langle\varepsilon\rangle .
\end{equation}
for every $\psi\in\mathcal{S}(\mathbf{R} )$.

In case when $p$ is equal to  $+\infty$  functions $u
(t,x,\varepsilon )$, $\rho (t,x,\varepsilon )$ and   $\sigma
(t,x,\varepsilon )$ exactly satisfies the system (\ref{system-2})
in the sense of
$\mathbf{R}\langle\varepsilon\rangle$--distributions.
\end{definition}

We will seek for a solution of this system in the  form
(\ref{velocity-function}), (\ref{stress-function}),
(\ref{velocity-expansion}), (\ref{stress-expansion}),
\begin{equation}\label{density-function}
\rho(t,x,\varepsilon)=\rho_0+\Delta \rho R
\left(\frac{x-vt}{\varepsilon}\right),
\end{equation}
$\rho_0, \Delta \rho, v$ are real numbers, $\Delta \rho\not=0$ and
$R(x)=\displaystyle\int\limits_{-\infty}^{x} \widetilde{R} (y) d
y$, $\displaystyle\int\limits_{-\infty}^{+\infty} \widetilde{R}
(y) d y=1$ and $\widetilde{R}\in \mathcal{S}(\mathbf{R} ).$ We
suppose
\begin{equation}\label{density-expansion}
\widetilde{R}(x)=b_0 h_0(x)+b_1 h_1(x)+\ldots+b_{n}
h_{n}(x),\,\,\, \vec{b}=(b_0, b_1, \ldots , b_n),
\end{equation}

Substituting (\ref{velocity-function}) and
(\ref{density-function}) into (\ref{solution-1ofthesystem-2}) we
get the following relations for the moments.

\begin{equation}\label{system-2conditions-1}
\{-v \Delta \rho + \Delta\rho u_0\} m_{k}(\widetilde{R}) +
\Delta\rho\Delta u m_k (\widetilde{R}U)+\rho_0\Delta u
m_k(\widetilde{U})+\Delta\rho\Delta u m_k (\widetilde{U}R)=0
\end{equation}
where $k=0,1,\ldots n.$. Note that
$m_0(\widetilde{R}U)=1-m_0(\widetilde{U}R)$. Suppose $k=0$ and we
get $$-v \Delta \rho + \Delta\rho u_0+\Delta\rho\Delta u
-\rho_0\Delta u=0 $$ or
\begin{equation}\label{velocity-value1}
v= u_0+\Delta u+\rho_0 \frac{\Delta u}{\Delta\rho}
\end{equation}
Last expression gives us the following
\begin{equation}\label{system-2conditions-1'}
-\Delta u (\rho_0 + \Delta\rho ) m_{k}(\widetilde{R}) +
\Delta\rho\Delta u m_k (\widetilde{R}U)+\rho_0\Delta u
m_k(\widetilde{U})+\Delta\rho\Delta u m_k (\widetilde{U}R)=0
\end{equation}
where $k=0,1,\ldots n.$.

All four vectors with coordinates $m_k (\widetilde{R})$, $m_k
(\widetilde{R}U)$, $m_k (\widetilde{U})$, $m_k (\widetilde{U}R)$
should be collinear. Consider $m_k (\widetilde{R})$ and $m_k
(\widetilde{U})$. Because of $m_0 (\widetilde{R}) = m_0
(\widetilde{U})=1$ then  $m_k (\widetilde{R}) = m_k
(\widetilde{U})$ for $k$ from $0$ to $n$ and therefore
$\vec{a}=\vec{b}$. From last equality we have

$$m_k (\widetilde{R}U)=m_k (\widetilde{U}R)=m_k (\widetilde{U}U)$$
where $k=0,1,\ldots n.$. Thus
\begin{equation}
-\Delta u (\rho_0 + \Delta\rho ) m_{k}(\widetilde{U}) +
2\Delta\rho\Delta u m_k (\widetilde{U}U)+\rho_0\Delta u
m_k(\widetilde{U})=0
\end{equation}
where $k=0,1,\ldots n.$. It means
\begin{equation}
m_{k}(\widetilde{U}) = 2 m_k (\widetilde{U}U), \,\,\,k=0,1,\ldots
n.
\end{equation}

Substituting (\ref{velocity-function}), (\ref{stress-function})
and (\ref{density-function}) into (\ref{solution-2ofthesystem-2})
we get the following relations for the moments.

\begin{equation}\label{system-2conditions-2}
\begin{array}{c}
u_0\Delta\rho (u_0-v) m_{k}(\widetilde{R}) + \rho_0\Delta u
(2u_0-v) m_k (\widetilde{U})+\\+\Delta u \Delta\rho (2u_0-v)\{
m_k(\widetilde{U}R)+ m_k(\widetilde{R}U)\}+\\+2\rho_0 (\Delta u)^2
m_k (\widetilde{U}U)+\rho (\Delta u)^2 m_k
(\widetilde{R}U^2)+\\+2\Delta\rho (\Delta u)^2 m_k
(\widetilde{U}RU)-\Delta\sigma m_k (\Sigma)=0
\end{array}
\end{equation}
where $k=0,1,\ldots n.$.
 Suppose
$k=0$ and we get
\begin{equation}\label{dencity-stress-velocity-value1}
\rho_0 (\Delta u)^2\left(\frac{\rho_0}{\Delta \rho}
+1\right)+\Delta\sigma =0.
\end{equation}
Moreover $m_{k}(\widetilde{R}) = m_k (\widetilde{\Sigma}),
\,\,\,k=0,1,\ldots n$ and then $\vec{b}=\vec{c}$. Finally,
$$\vec{a}=\vec{b}=\vec{c}\,\,\, \textrm{and} \,\,\,
m_{k}(\widetilde{R}) =3 m_k (\widetilde{U}U^2), \,\,\,k=0,1,\ldots
n.$$

Substituting (\ref{velocity-function}) and (\ref{stress-function})
into (\ref{solution-3ofthesystem-2}) we get the following
relations for the moments.

\begin{equation}\label{system-2conditions-3}
\{u_0\Delta\sigma -v \Delta\sigma\} m_{k}(\widetilde{\Sigma}) +
\Delta u \Delta\sigma m_k (\widetilde{\Sigma}U)-k^2\Delta u
m_k(\widetilde{U})=0
\end{equation}
where $k=0,1,\ldots n.$.
 Suppose
$k=0$ and using expression for the velocity
(\ref{velocity-value1}) we get
\begin{equation}\label{dencity-stress-velocity-value2}
\Delta\sigma\Delta u \left(\frac{\rho_0}{\Delta \rho}
+\frac{1}{2}\right)+k^2\Delta u =0.
\end{equation}
From the (\ref{system-2conditions-3}) we can also find the
following equality for the velocity
\begin{equation}\label{velocity-value2}
v= u_0+\Delta u+\frac{1}{2} u_0-k^2 \frac{\Delta u}{\Delta\sigma}
\end{equation}
It is well known result in the elasticity theory.

Thus, if $\rho_0$, $\Delta u$ and $k^2$ are known then the rest
constants we can find from the system

\begin{equation}\label{dencity-stress-velocity-system}
\left\{\begin{array}{c} \rho_0 (\Delta u)^2\left(\rho_0+\Delta
\rho\right)+\Delta\sigma\Delta\rho =0, \\
\Delta\sigma\left(\rho_0+\frac{1}{2}\Delta
\rho\right)+k^2\Delta\rho =0. \end{array}\right.
\end{equation}
Hence,
\begin{equation}
\Delta\sigma=-\frac{2k^2\Delta\rho}{2\rho_0+\Delta\rho}
\end{equation}
and
\begin{equation}
\Delta\rho_{1,2}=\frac{-\frac{3}{2}\rho_0^2(\Delta
u)^2\pm\sqrt{\frac{\rho_0^4 (\Delta u)^4}{4}+4k^2(\Delta
u)^2\rho_0^3}}{\rho_0 (\Delta u)^2-2k^2}.
\end{equation}

\begin{theorem}\label{theorem5}
For any integer $p$ there is a solution of the system of equations
(\ref{system-2}) in the sense of the definition 6. Moreover,
$$\left\{\begin{array}{l} \rho_0 (\Delta u)^2\left(\rho_0+\Delta
\rho\right)+\Delta\sigma\Delta\rho =0, \\
\Delta\sigma\left(\rho_0+\frac{1}{2}\Delta
\rho\right)+k^2\Delta\rho =0,\\v= u_0+\Delta u+\rho_0 \frac{\Delta
u}{\Delta\rho}. \end{array}\right.$$
\end{theorem}

Shock profiles of the considered system (\ref{system-2}) can be
found on pictures (Fig. \ref{fig21}, \ref{fig22}, \ref{fig23},
\ref{fig24}). We considered case when $p=13$ and $\Delta u=-1$,
$u_0=1$,  $k^2=0.1$. We can also take any real $\rho_0, \sigma_0$
(see conditions of the theorem \ref{theorem5}) but here we took
$\rho_0=1.1$, $\sigma_0=0.5$ and then we should calculate
$\Delta\rho$ and $\Delta\sigma$. We consider two different types
of shock profiles. The first is on the  Fig. \ref{fig21},
\ref{fig22}. The second is on the  Fig. \ref{fig23}, \ref{fig24}.

\begin{figure}[ht]
\begin{minipage}[b]{.49\linewidth}
\centering\includegraphics[width=\linewidth]{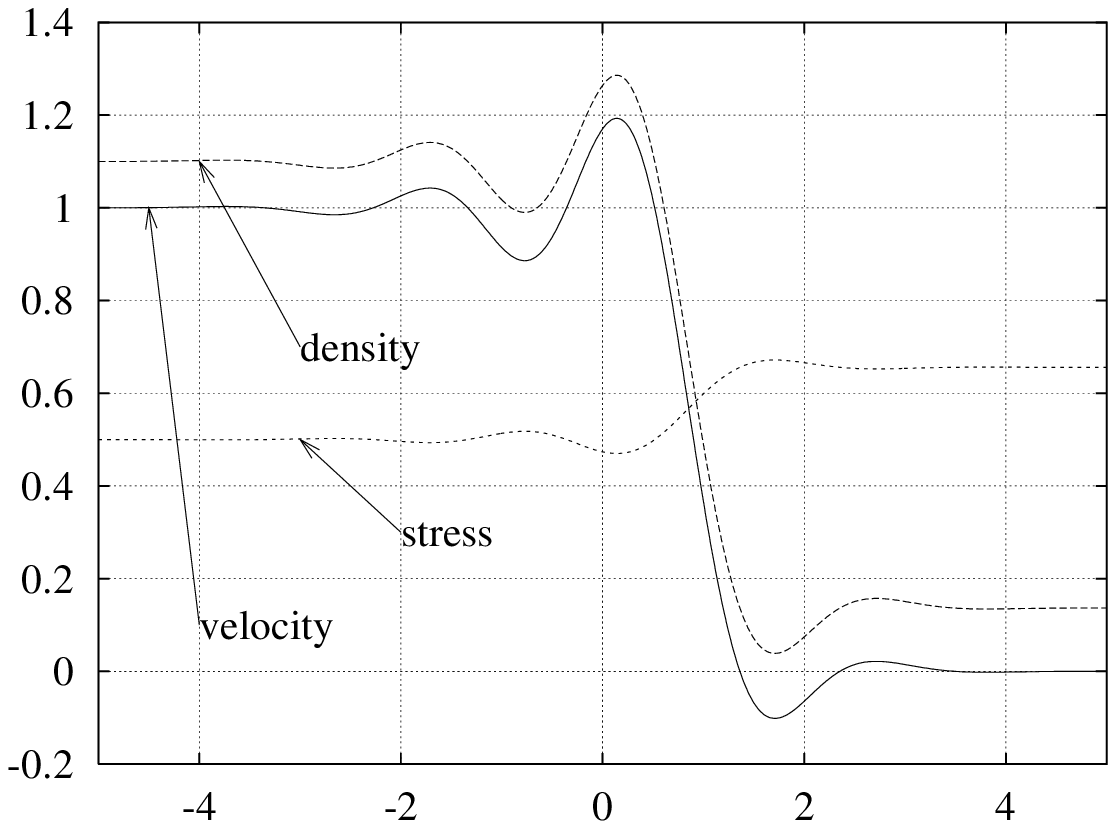}
\caption{\footnotesize Shock profiles, $v=1.1417$.}\label{fig21}
\end{minipage}\hfill
\begin{minipage}[b]{.49\linewidth}
\centering\includegraphics[width=\linewidth]{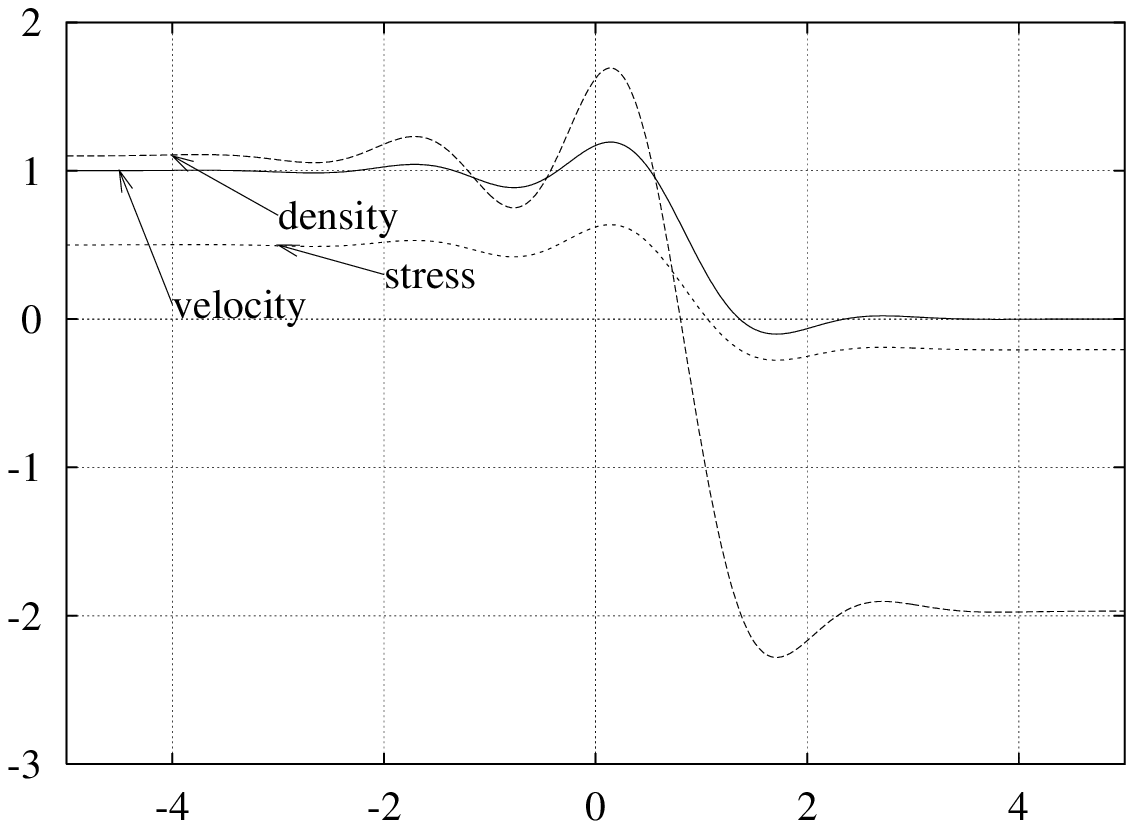}
\caption{\footnotesize Shock profiles, $v=0.35833$.}\label{fig22}
\end{minipage}
\end{figure}

\begin{figure}[ht]
\begin{minipage}[b]{.49\linewidth}
\centering\includegraphics[width=\linewidth]{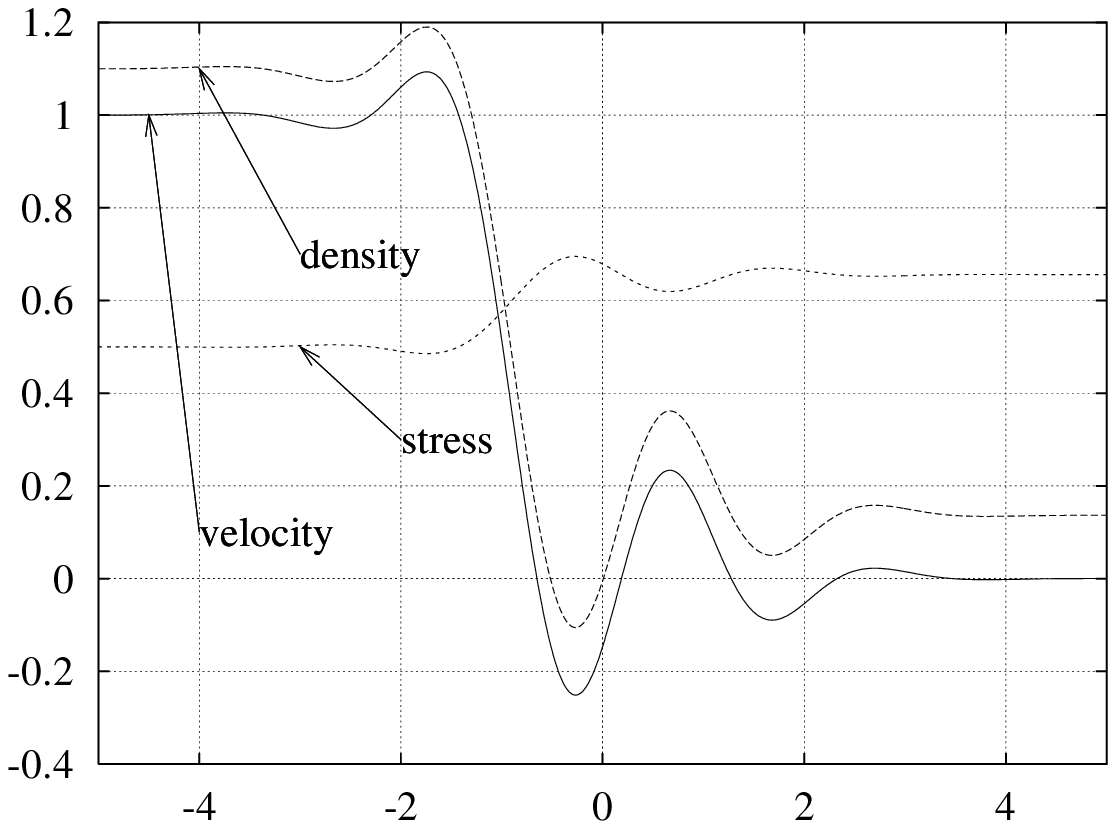}
\caption{\footnotesize Shock profiles, $v=1.1417$.}\label{fig23}
\end{minipage}\hfill
\begin{minipage}[b]{.49\linewidth}
\centering\includegraphics[width=\linewidth]{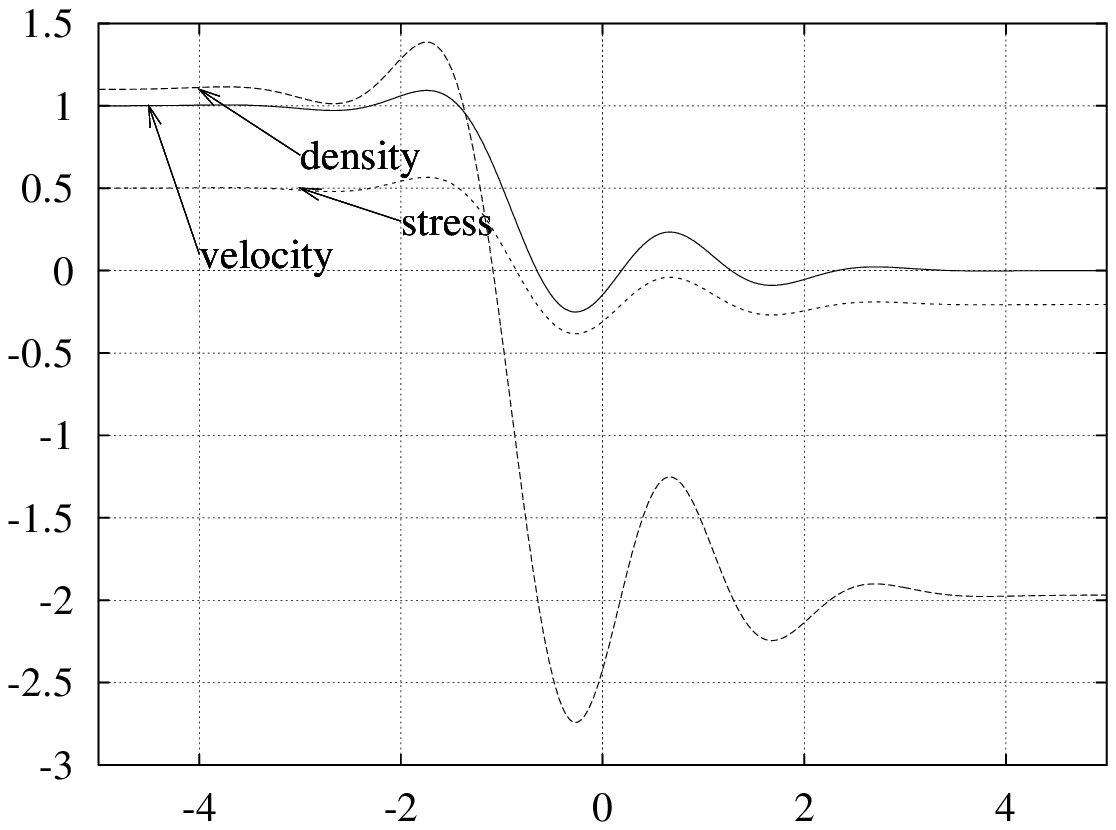}
\caption{\footnotesize Shock profiles, $v=0.35833$.}\label{fig24}
\end{minipage}
\end{figure}

\section{Conclusions and remarks.} In fact, we considered only special kind
of solutions from the sets $I$ and $J$. Moreover, mentioned
solutions are ``approximate'' solutions. It is open question about
existence of the solution of the Hopf equation in sense of the
definition \ref{definition4} when $p=\infty$.

We should notice that there is also a Non-Archimedean approach
which is developed by V.Vladi\-mirov, I.Volovich, E.Zelenov
\cite{Vladimirov-Volovich-Zelenov}, A.Khrennikov
\cite{Khrennikov}. This approach based on $p$-adic valued
distributions and  used for the construction of some models in
Mathematical Physics.

The authors of the papers \cite{Egorov}, \cite{Colombeau-LeRoux}
consider the same equations but they speculated a different
ideology for generalized solutions and generalized functions.

In conclusion we should emphasis that our calculation method looks
like the Fourier method for linear differential equations but
applied to the nonlinear equations. Compare the method of mode
superposition for a string and our method for the shock. Our
method allowed to obtain all known formulas for the shocks
characteristics and, in addition, find a microscopic behaviour of
shocks in the thin layer with an assumption that the profile of
the shock can be approximated by the orthogonal system of
functions. We can use Laguerre functions, harmonic functions or
any orthogonal system in our calculations instead of Hermite
functions.

Our method one can apply to the problems of hydrodynamics, quantum
mechanics and non-linear optics.

\textbf{Acknowledgment}

It is pleasure to thank the the seminar of Moscow Energy Institute
organized by Prof. Yu.A. Dubinskii for the consideration to this
work.

Research is partially supported by Belarussian Fundamental
Research Foundation Grant No F99M-082.

\end{document}